\documentclass[twocolumn]{aastex631}
\usepackage{ae,aecompl,multirow}
\usepackage{threeparttable}
\usepackage{booktabs}
\usepackage{url}
\usepackage{amsmath}
\usepackage{graphics}

\begin{document}

\title{Spectrophotometric reverberation mapping of Intermediate-mass black hole NGC 4395}

\correspondingauthor{Shivangi Pandey}
\email{Shivangipandey@aries.res.in}

\author[0000-0002-4684-3404]{Shivangi Pandey}
\affiliation{Aryabhatta Research Institute of Observational Sciences, Nainital\textendash263001, Uttarakhand, India}
\affiliation{Department of Applied Physics/Physics, Mahatma Jyotiba Phule Rohilkhand University, Bareilly\textendash243006, India}

\author[0000-0002-8377-9667]{Suvendu Rakshit}
\affiliation{Aryabhatta Research Institute of Observational Sciences, Nainital\textendash263001, Uttarakhand, India}
\email{suvenduat@gmail.com}

\author[0000-0002-6789-1624]{Krishan Chand}
\affiliation{Department of Physics and Astronomical Science, Central University of Himachal Pradesh, Dharamshala, Kangra\textendash176215, Himachal Pradesh, India}

\author[0000-0002-4998-1861]{C. S. Stalin}
\affiliation{Indian Institute of Astrophysics, Block II, Koramangala, Bangalore\textendash560034, India}

\author[0000-0003-2010-8521]{Hojin Cho}
\affiliation{Department of Physics \& Astronomy, Seoul National University, Seoul\textendash08826, Republic of Korea}

\author{Jong-Hak Woo}
\affiliation{Department of Physics \& Astronomy, Seoul National University, Seoul\textendash08826, Republic of Korea}

\author{Priyanka Jalan}
\affiliation{Center for Theoretical Physics of the Polish Academy of Sciences, Al. Lotnik$\Acute{o}$w 32/46, 02-668 Warsaw, Poland}

\author{Amit Kumar Mandal}
\affiliation{Center for Theoretical Physics of the Polish Academy of Sciences, Al. Lotnik$\Acute{o}$w 32/46, 02-668 Warsaw, Poland}
\affiliation{Astronomy Program, Department of Physics and Astronomy, Seoul National University, Seoul\textendash08826, Republic of Korea}

\author{Amitesh Omar}
\affiliation{Indian Institute of Technology Kanpur, Kanpur\textendash208016, Uttar Pradesh, India}

\author{Jincen Jose}
\affiliation{Aryabhatta Research Institute of Observational Sciences, Nainital\textendash263001, Uttarakhand, India}
\affiliation{Center for Basic Sciences, Pt. Ravishankar Shukla University, Raipur, Chhattisgarh\textendash492010, India}

\author{Archana Gupta}
\affiliation{Department of Applied Physics/Physics, Mahatma Jyotiba Phule Rohilkhand University, Bareilly\textendash243006, India}

\begin{abstract}
Understanding the origins of massive black hole seeds and their co-evolution with their host galaxy requires studying intermediate-mass black holes (IMBHs) and estimating their mass. However, measuring the mass of these IMBHs is challenging due to the high spatial resolution requirement. A spectrophotometric reverberation monitoring is performed for a low-luminosity Seyfert 1 galaxy NGC 4395 to measure the size of the broad line region (BLR) and black hole mass. The data were collected using the 1.3-m Devasthal fast optical telescope (DFOT) and 3.6-m Devasthal optical telescope (DOT) at ARIES, Nainital, over two consecutive days in March 2022. The analysis revealed strong emission lines in the spectra and light curves of merged 5100{\AA} spectroscopic continuum flux ($f_{\mathrm{5100}}$) with photometric continuum V-band and H$\alpha$, with fractional variabilities of 6.38\% and 6.31\% respectively. In comparison to several previous studies with lag estimation $<$ 90 minutes, our calculated H$\alpha$ lag supersedes by $125.0^{+6.2}_{-6.1}$ minutes using ICCF and {\small JAVELIN} methods. The velocity dispersion ($\sigma_{\mathrm{line}}$) of the broad line clouds is measured to be $544.7^{+22.4}_{-25.1}$ km s$^{-1}$, yielding a black hole mass of $\sim$ $2.2^{+0.2}_{-0.2}\times 10^{4}M_{\mathrm{\odot}}$ and an Eddington ratio of 0.06. 
\end{abstract}

\section{Introduction}
\label{sec:intro}
Understanding the growth and evolution of supermassive black holes (SMBHs) formed in the early Universe is one of the most fundamental problems. Recent observations of JWST suggest the presence of massive black holes at z$>$7, challenging the theories of black hole formation and evolution at early Universe \citep[e.g.,][]{2023ApJ...957L...7K}.  
The origin of these massive seeds is believed to be heavier than stellar mass black holes of mass ranging in $\sim$ 10-100M$_{\mathrm{\odot}}$. Hence, their progenitor is suggested to be intermediate-mass black holes (IMBHs) with masses 10$^{3}$ - 10$^{6}$M$_{\mathrm{\odot}}$, making them currently active research subjects \citep[e.g.,][]{loeb_collapse_1994, latif_characteristic_2013, greene_intermediate-mass_2020}. However, the detection of IMBHs is challenging mainly due to their low luminosity and negligible variability \citep[see][]{mezcua_observational_2017, greene_intermediate-mass_2020}, and very few candidates have been known so far \citep[e.g.,][]{greene_active_2004, greene_new_2007, reines_dwarf_2013, Shin_2022, Ward2022, Yang2022, Samantha2024}.

The study of the mass distribution and co-evolution of black holes and their host galaxies is enabled by the correlation between the black hole masses and the stellar velocity dispersion $\sigma_{\mathrm{\star}}$ of the bulge of the galaxy \citep[see,][]{Jian2008, Woo2015} as identified by \citet{kormendy_coevolution_2013}. It is beyond the capability of modern instruments to spatially resolve the central engine of AGNs \citep[however, see recent results from][]{Gravity2018, GRAVITYCollaboration2020, GRAVITY2021, Abuter_2024}, hampering accurate measurement of their black hole masses. Consequently, most of the studies on the central engine of AGN are based on reverberation mapping \citep[RM;][]{Peterson_1993}. RM observes the response of emission lines to continuum variations from the central source, thus studying the size and structure of the broad-line region (BLR). This method has been applied to more than 100 objects \citep[e.g.,][]{Peterson_1993, Peterson_1998, Kaspi2000, Kaspi2007, Peterson_2002, Peterson_2004, Dietrich2012, bentz_low-luminosity_2013, Peterson2014, Du2014, Du2015, Woo2015, Du_2016, Pei2017, Fausnaugh2017, Grier2017, Park_2017, Du_2018, Rakshit2019, Bonta2020, Rakshit2020A&A, Williams2020, Cackett2021, Hu2021, Li2021, U2022,pandey_spectroscopic_2022, Malik2023, Woo2024}. Unfortunately, only very few RM-based studies have been conducted for IMBHs due to low variability \citep[see,][]{peterson_multiwavelength_2005, Rafter2011,cho2021}.

NGC 4395 is a low luminosity Seyfert 1 galaxy with strong emission lines at a redshift of 0.001, hosting an IMBH candidate \citep{filippenko_discovery_1989}. It contains the least luminous broad-lined AGN known to date, with a bolometric luminosity lower than 10$^{41}$ erg s$^{-1}$ and stellar mass $\sim$ 10$^{9}$ M$_{\mathrm{\odot}}$ \citep{filippenko_discovery_1989, filippenko_low-mass_2003, cho2021}. The Eddington ratio of NGC 4395 is $\sim$ 5$\%$ \citep{woo_10000-solar-mass_2019}, comparable to that of other RM AGNs. Therefore, this galaxy provides a unique opportunity to investigate photoionization and size-luminosity relations at the extreme low-luminosity regime of AGN. The exact mass of central black hole of NGC 4395 is in conflict, with estimates ranging from 9 $\times$ 10$^{3}$ M$_{\mathrm{\odot}}$ to 4 $\times$ 10$^{5}$ M$_{\mathrm{\odot}}$ \citep[e.g.,][]{filippenko_low-mass_2003, peterson_multiwavelength_2005, edri_broadband_2012, woo_10000-solar-mass_2019}. 

Measuring an emission-line lag for NGC 4395 has been challenging due to low continuum variation and the variable component being weak for emission lines such as HeII, H$\gamma$, and mostly for H$\beta$. However, various attempts have been made for lag measurement such as for H$\alpha$ \citep{Desroches2006, Edri2012, woo2019, Cho_2020, cho2021}, for H$\beta$ \citep{Edri2012}, for Pa$\beta$ \citep{Franca2015}, for CIV \citep{peterson_multiwavelength_2005}. \citet{woo_10000-solar-mass_2019} provided the first reliable H$\alpha$ lag of 83$\pm$14 minutes using photometric RM (PRM), which, combined with the line width of 426$\pm$2.5 km s$^{-1}$ from a single-epoch spectrum leads to a black hole mass estimate of 10,000 M$_{\mathrm{\odot}}$. However, the photometric H$\alpha$ filter has a continuum contribution that can be a significant source of uncertainty in estimating a true lag. Therefore, quantifying the continuum variability and contribution in the H$\alpha$ filter is crucial. A further PRM attempt was made by \citet{Cho_2020} through extensive photometric monitoring campaigns where the authors employed various scaling parameter values to correct the continuum contribution in the H$\alpha$ narrow-band filter.
The lag value is found to have a large range, from 55 to 122 minutes, when the H$\alpha$ narrow-band flux was corrected for continuum, assuming this contribution to be 0 to 100 \% of V-band variability. In spectroscopic RM, such correction is not needed as flux-calibrated spectra provide the continuum variation, which can be measured by modeling it with a power-law alongside decomposed emission lines. Therefore, spectroscopic RM is crucial to estimate BLR size and calibrate the size-luminosity relation of the extremely low-luminosity region and accurate black hole mass.

Previous attempts of spectroscopic RM have either been inconclusive or provided a lag upper limit. \citet{cho2021} made a noteworthy attempt using high-cadence imaging data for NGC 4395. As a follow-up of previous PRM \citep{woo_10000-solar-mass_2019, Cho_2020}, they provided an improved H$\alpha$ line dispersion measurement. However, due to the lack of variability in the light curve and bad weather losses, only a lag upper limit of less than 3 hours has been provided. To measure BLR size and black hole mass using spectroscopic RM of this low luminous Seyfert 1 galaxy NGC 4395, we conducted a comprehensive photometric and spectroscopic monitoring program for two consecutive nights. In this paper, we present the result of this monitoring program measuring the BLR size and black hole mass.

The paper is organized as follows. The observations and data reduction have been discussed in section \ref{sec:Observation}. The results and analysis have been presented in section \ref{sec:Results}. The measurement of black hole mass and a few important aspects are discussed in section \ref{sec:Discussion} and finally concluded in section \ref{sec:Conclusions}.

\section{Observations and Data Reduction}
\label{sec:Observation}
NGC 4395 was observed using two telescopes hosted and operated by Aryabhatta Research Institute of Observational Sciences, Nainital, India (ARIES) from 10$^{\mathrm{th}}$ to 11$^{\mathrm{th}}$ March 2022, with observations spanning 7-8 hours on both nights, which exceeds the expected lag of the source by multiple times. 

I) Photometric observations were conducted using the 1.3-m Devasthal fast optical telescope \citep[DFOT;][]{sagar_new_2011} equipped with a 2k $\times$ 2k CCD Camera providing a plate scale of 0$^{\mathrm{\prime \prime}}$.53/pixel. The source NGC 4395 was observed in the V-band and narrow-band filters H$\alpha$ and [\ion{S}{2}], with exposures of 300 sec in the sequence V-H$\alpha$-[\ion{S}{2}]. Consequently, the source was observed in all three filters with a 10-minute cadence. Approximately 42 photometric data points were obtained in the V-band. In this work, the V-band photometry results are presented. Photometry RM results using narrow-band H$\alpha$ and [\ion{S}{2}] data will be reported elsewhere.

II) Spectroscopy of the source was performed using the 3.6-m Devasthal optical telescope \citep[DOT;][]{Kumar2018}. The ARIES Devasthal-Faint Object Spectrograph and Camera \citep[ADFOSC;][]{omar_first-light_2019} mounted at the 3.6-m DOT were used for spectroscopic observations. It is comprised of a 4k $\times$ 4k deep depletion CCD Camera providing a 0$^{\mathrm{\prime \prime}}$.2/pixel scale with 2 $\times$ 2 binning \citep{Dimple2022}. The spectroscopic observation was carried out with a slit 3$^{\mathrm{\prime \prime}}$.2 wide and 8 $^{\mathrm{\prime \prime}}$ long and a 132R$-$600 gr/mm grism that covers the wavelength range of 3500-7000{\AA}, centered at 4880{\AA}. Spectroscopic frames of 300 sec duration were taken throughout the night, along with bias and flat frames for pre-processing. The seeing varied throughout the night, ranging from 0$^{\mathrm{\prime \prime}}$.5 to 1$^{\mathrm{\prime \prime}}$.5. The instrumental resolution is measured to be 7{\AA} (312 km s$^{-1}$) by modeling the emission lines present in the combined Hg-Ar Ne lamp obtained in the same configuration as the source frame and also verified using skylines present in the observed spectrum of NGC 4395.

\begin{figure}
\centering
\includegraphics[width=8cm, height=7cm]{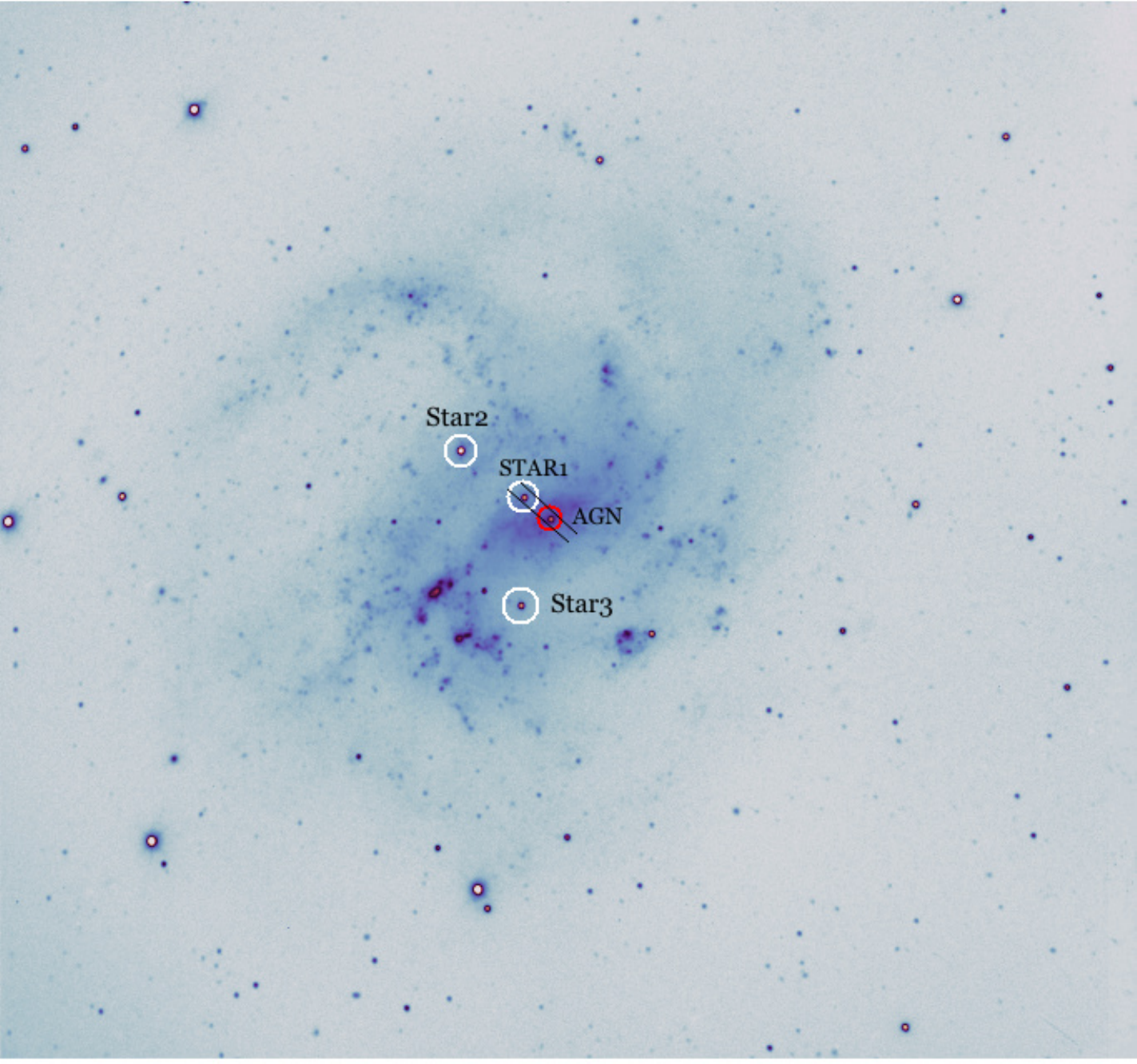}
\caption{The combined V-band image of NGC 4395 obtained on 10$^{\mathrm{th}}$ March 2022 using 1.3-m DFOT. Three comparison stars are marked with white circles, while the target AGN is denoted with a red circle. The slit is oriented to cover both the AGN and a comparison star. The field of view is 66 arcmin (18$^{\prime}$ $\times$ 18$^{\prime}$) for 1.3-m DFOT.}
\label{fig:imagengc}
\end{figure}

Detecting variability in the central engine of the low-luminosity NGC 4395 is difficult due to the contribution from its extended host galaxy and narrow line region (NLR). Considering the narrow line region (NLR) is spatially resolved using ground-based telescopes, this leads to variable contributions in the slit spectra, which makes the standard approach of relative flux calibration using narrow lines impractical \citep[see][]{cho2021},  making accurate flux calibration with respect to a reference star essential. For this purpose, the slit was oriented to encompass a nearby comparison star, as depicted in Fig. \ref{fig:imagengc}. This allowed for the simultaneous acquisition of spectra for both the source and a steady comparison star. The number of photometric data points and spectra obtained from both telescopes are detailed in Table \ref{tab:observation data}. Approximately 100 spectra were obtained from the 3.6-m DOT, sufficient for lag measurement through time series analysis.
 
\begin{table}
\caption{Photometric Observations using 1.3-m DFOT in V-band filter and Spectroscopic observations from 3.6-m DOT (ADFOSC) on 10$^{\mathrm{th}}$ and 11$^{\mathrm{th}}$ March, 2022.}
\begin{tabular}{ccc}
\hline
Observation & \multicolumn{2}{c}{Data} \\ \cline{2-3}
&Photometric V-band& Spectra\\
& (1.3-m DFOT) &(3.6-m DOT) \\
\hline
10$^{\mathrm{th}}$ March & 19 & 43 \\
11$^{\mathrm{th}}$ March & 23 & 64 \\
\hline
\end{tabular}
\begin{tablenotes}
\item \textbf{Note.} Columns: Observation date, number of data points observed from 1.3-m DFOT, and spectral data obtained from 3.6-m DOT.
\end{tablenotes}
\label{tab:observation data}
\end{table}

\subsection{Differential Photometry}
\label{sec:Photometry}
The NGC 4395 is an extended spiral galaxy with the core marked with red circles as shown in Fig. \ref{fig:imagengc} and the three steady comparison stars marked with white circles. These comparison stars are used to perform differential photometry. Initially, the images from respective days were aligned using a Python package named Astroalign \citep{Astroalign}. The alignment ensured that the point spread function (PSF) matched that of the image with the lowest resolution, resulting in frames with uniform PSF. The pre-processing stage involved cleaning the photometric frames, i.e., bias-subtraction, flat-fielding and cosmic ray correction. Aperture photometry was carried out using a Python wrapper called SEP \citep{Barbary2016} for source extraction.

The aperture size was $2.5$ times the average full width at half maximum (FWHM) of all three comparison stars. The FWHM was calculated by fitting the Gaussian function. The sky contribution was assumed to be 4 to 5 times the FWHM. The differential magnitude of the source compared to the three comparison stars present in the same field was then calculated. Finally, the zero point was added to convert the differential instrumental magnitude into actual V-band magnitude. The light curve obtained from the 1.3-m DFOT in the V-band on both days is depicted in Figure \ref{fig:lightcurve_V}. The light curve is merged with the continuum flux at 5100{\AA} ($f_{\mathrm{5100}}$) obtained from the spectra, as discussed in section \ref{sec:Spectroscopy}.

\begin{figure*}
\includegraphics[height=9cm, width=18cm]{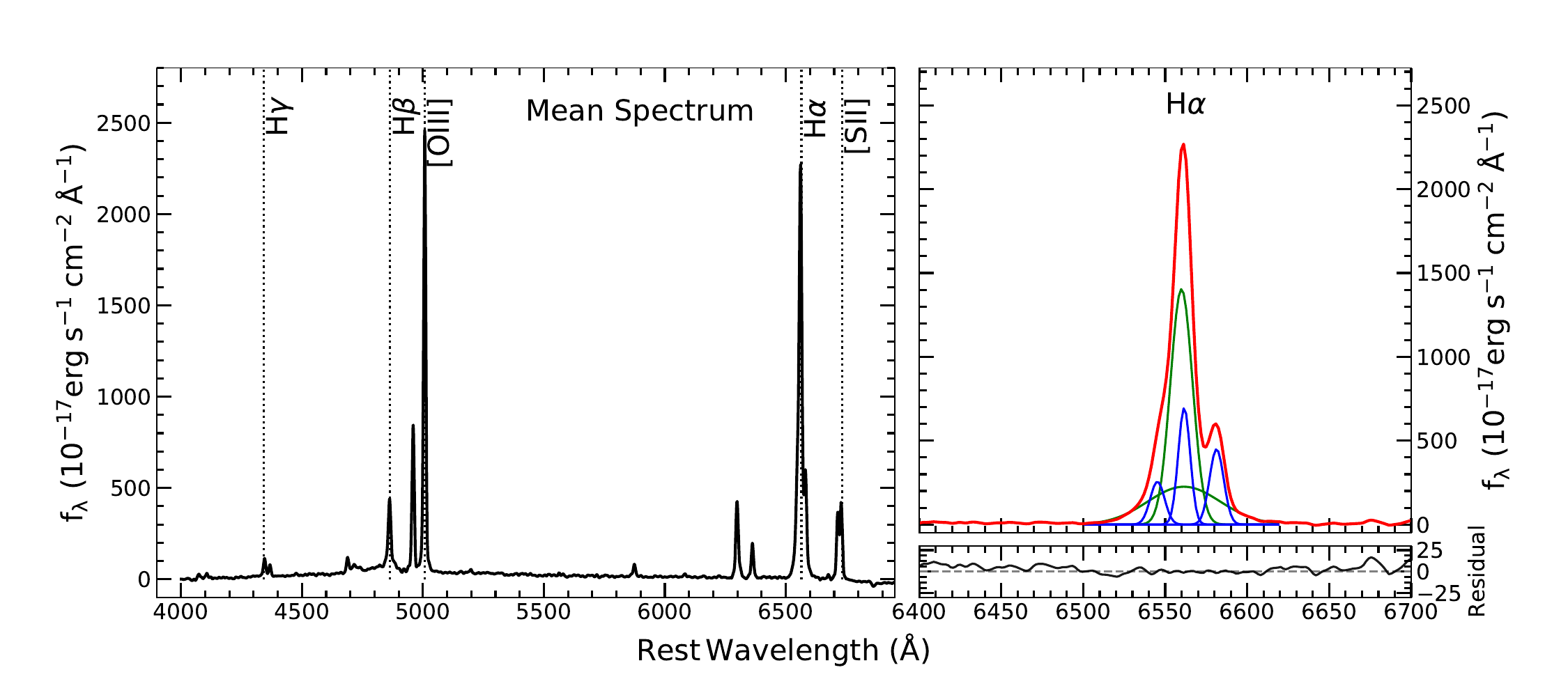}
\caption{An example of spectral decomposition of the mean spectrum of 11$^{\mathrm{th}}$ March. The left panel shows the mean spectrum with marked emission line regions. The right panel presents the decomposition of the H$\alpha$ component into narrow and broad components with narrow [\ion{N}{2}] lines. The broad H$\alpha$ component is modeled with the double Gaussian shown in green, and the narrow component is with the single Gaussian shown in blue. The narrow [\ion{N}{2}] $\lambda6549$ and $\lambda6585$ emission lines are shown with a single Gaussian fit in blue. The total line construction is shown in red with the illustrated continuum region. The lower right panel shows the residual of the fit in black.}
\label{fig:Mean-Spectrum}
\end{figure*}

\subsection{Spectroscopic Reduction and flux calibration}
\label{sec:Spectroscopy}
\subsubsection{Spectroscopy}
The spectroscopic reduction was performed using IRAF software \citep{Tody1986, Tody1993, IRAF}, encompassing bias subtraction, flat fielding, and cosmic ray removal with the L.A. cosmic algorithm \citep{dokkum_cosmicray_2001}. Flat correction was executed using a Tungsten-LED lamp, which was normalized. The bias-corrected frames, including the source and lamps, were divided by it. The apertures of the center of NGC 4395 and comparison stars are extracted with a size of 4 $^{\mathrm{\prime \prime}}$ using task `apall', which is included in the IRAF software. Additionally, HgAr and Neon lamps were observed in the same configuration as the source frame for wavelength calibration covering the wavelength range of 3500-7000{\AA}. Both of these lamps were combined using the `imcombine' task. The combined lamps were identified, and the final wavelength solution was applied to the source and reference star frames to complete the wavelength calibration process.

\subsubsection{Flux calibration}
\label{sec:flux calibration}
Accurate flux calibration is imperative for studying emission line variability. In RM studies, narrow-line fluxes are typically relied upon for precise flux calibration, assuming that the narrow-line fluxes remain constant throughout the monitoring campaign. However, this assumption does not hold for NGC 4395, having extended and resolved host galaxy and NLR. The flux contamination, in this case, depends upon the seeing variation and its variability throughout the monitoring campaign. Moreover, \citet{cho2021} have identified significant variability from the narrow [\ion{S}{2}] line fluxes in the slit-based spectroscopy. This renders the use of narrow-line fluxes for relative flux calibration unsuitable.

We follow the procedure outlined in \citet{cho2021} for flux calibration. As the in-slit comparison star S1 is not a standard star, its slope and features were matched with HD 165341 (K0 V) (hereafter $L1_{\mathrm{cat}}$) from the Indo-US Library \citep[similar to,][]{cho2021}. The library spectrum $S_{\mathrm{lib}}$ was scaled to match the known magnitude of S1 from the \href{https://vizier.cds.unistra.fr/viz-bin/VizieR-S?2MASS%20J12255090%2b3333100}{Vizier Online Data Catalog} \citep{2020yCat.1350....0G}. Subsequently, the scaled library spectrum ($S_{\mathrm{lib,scaled}}$) was used to create a response function for flux calibration. To generate the response curve ($S_{\mathrm{\star}}$), a polynomial function was fitted by considering the line-free region of the scaled library spectrum $S_{\mathrm{lib,scaled}}$. Similarly, this process was repeated for star S1 epoch by epoch, denoted as $S1_i$. The sensitivity curve $S_{\mathrm{i}}$ was then calculated as $S_{\mathrm{i}}=S_{\mathrm{\star}}$/$S1_i$ for the $i$-th epoch. Finally, the source wavelength-calibrated spectra were multiplied by the sensitivity function $S_{\mathrm{i}}$, resulting in its flux calibration.

Finally, the flux-calibrated spectra obtained from the method above were recalibrated with V-band photometric data points acquired using the 1.3-m DFOT. Subsequently, the V-band magnitude was converted to flux density using the formula V = 15 $-$ 2.5 log (F$_{\mathrm{\nu}}$/3.64). This photometric V-band flux density was used to scale the integrated region 5400-5600{\AA} of the spectra, resulting in the final flux-calibrated spectra.

\subsubsection{Spectral decomposition}
\label{sec:decomp}
The publicly available multi-component spectral fitting code PyQSOFit was used for decomposition and spectrum fitting, developed by \cite{GuoPyqsofit} and \citet{guo_legolasonpyqsofit_2023}. A comprehensive description of the code and its applications can be found in \citet{guo_constraining_2019}, \citet{shen_sloan_2019} and  \citet{Rakshit2020ApJS}. Each AGN spectrum underwent correction for Galactic extinction using the map by \cite{Schlegel1998} and the Milky Way extinction law of \cite{Fitzpatrick1999} with $R_V=3.1$. Subsequently, de-redshifting was performed using the redshift (z=0.001). Following this, the continuum was modeled using a power law in the line-free region of the spectrum and Fe II templates using \cite{Boroson1992}. After subtracting the continuum, detailed multi-Gaussian modeling was conducted in the H$\alpha$ region to fit the emission lines shown in Fig. \ref{fig:Mean-Spectrum}. The Narrow H$\alpha$, [\ion{N}{2}]$\lambda$6549, [\ion{N}{2}]$\lambda$6585, [\ion{S}{2}]$\lambda$6718, and [\ion{S}{2}]$\lambda$6732 lines were modeled using a single Gaussian with the velocity and velocity offset tied to each other. The broad H$\alpha$ component was modeled with double Gaussians. The best-fit model was obtained via $\chi^2$ minimization. Subsequently, the emission line flux, width, and continuum luminosity at 5100{\AA} with a window of 20{\AA} on both sides were calculated from the best-fitting model.

Fig. \ref{fig:Mean-Spectrum} presents an example of spectral fitting with PyQSOFit, where various emission lines are delineated, including broad and narrow emission line components.
Since the broad component of H$\alpha$ emanates from the broad emission line region (BLR), the broad H$\alpha$ flux must be used to construct the emission line light curve. However, due to the low-medium S/N of spectra (i.e., S/N around 10 to 20 at continuum 6100-6200{\AA}), constraining the line wing is difficult as it is prone to S/N. Hence, we used the best-fit broad H$\alpha$ model profile obtained from the PyQSOFit fitting and integrated it in the wavelength range of 6500-6600{\AA} to create the H$\alpha$ emission line light curve, avoiding wings. 

\subsection{Generating Lightcurve}
Due to the limited number of photometric data points on both nights, the spectroscopic flux at 5100{\AA} was used to augment the cadence in the continuum light curve. To mitigate systematic deviations introduced by merging two continuum light curves observed from different telescopes, both the photometric and $f_{\mathrm{5100}}$ light curves were inter-calibrated using the Python module PyCALI\footnote{\url{https://github.com/LiyrAstroph/PyCALI}}, as described in \citet{Li2014}. PyCALI employs a damped random walk (DRW) with a Bayesian approach to model the inter-calibrated light curves and estimates parameters such as the amplitude of variation ($\sigma$), time scale ($\tau$), scale, and shift factors for inter-calibrating the light curves.

\begin{figure*}
    \centering
    \includegraphics[height=10cm,width=17cm]{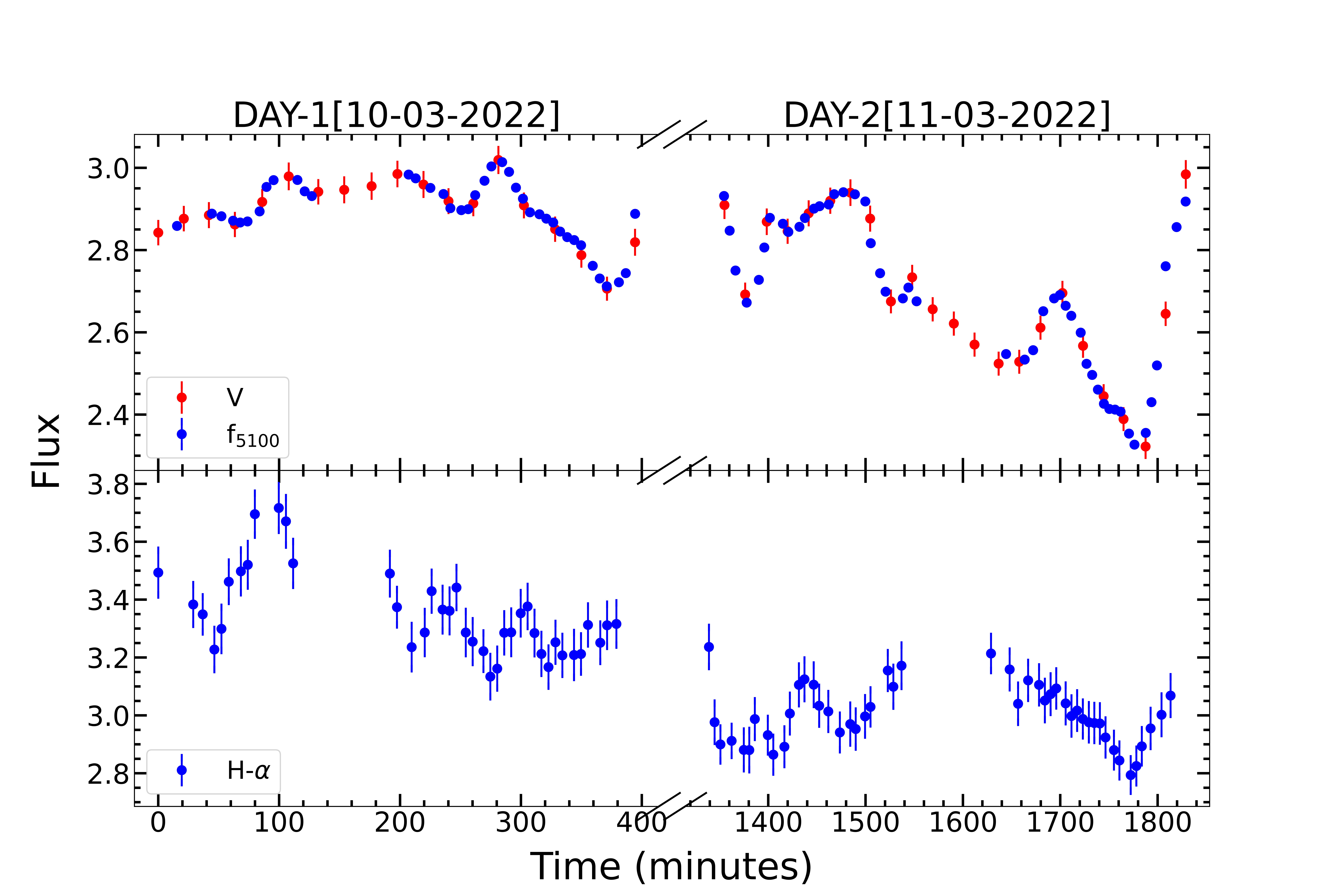}
    \caption{The upper panel is the merged light curve of photometric V-band and spectroscopic $f_{\mathrm{5100}}$ in units of 10$^{\mathrm{-15}}$ erg cm$^{\mathrm{-2}}$s$^{\mathrm{-1}}$\AA$^{\mathrm{-1}}$ and the lower panel is the H$\alpha$ emission line light curve in units of 10$^{\mathrm{-14}}$ erg cm$^{\mathrm{-2}}$s$^{\mathrm{-1}}$ obtained from spectroscopic data after smoothing by five consecutive data points.}
\label{fig:lightcurve_V}
\end{figure*}

The upper panel in Fig. \ref{fig:lightcurve_V} illustrates the merged V-band photometric flux density and $f_{\mathrm{5100}}$ light curves. Additionally, the lower panel displays the H$\alpha$ emission line light curve on both nights. To mitigate noise, particularly in the H$\alpha$ light curve, these light curves were smoothed by binning five continuous data points. 

\begin{table*}
\caption{\textbf{Light curve table}}
\centering
\begin{tabular}{lrrrrr}
\hline
\textbf{Time\_MJD} & \textbf{c\_flux} & \textbf{c\_error} & \textbf{l\_flux} & \textbf{l\_error} & \textbf{Type} \\
(1) & (2) & (3) & (4) & (5) & (6) \\
\hline
59648.953264 & 2.8425 & 0.0217 & -        & -        & P \\
59648.964006 & 2.8587 & 0.0091 & 34.9334  & 0.7723   & S \\
59648.967928 & 2.8765 & 0.0220 & -        & -        & P \\
59648.982326 & 2.8850 & 0.0226 & -        & -        & P \\
59648.984049 & 2.8886 & 0.0093 & 33.8328  & 0.6847   & S \\
59648.989542 & 2.8823 & 0.0091 & 33.4908  & 0.6045   & S \\
\hline
\end{tabular}
\begin{tablenotes}
\item \textbf{Note.} The columns are as follows: (1) Modified Julian Date (Time\_MJD); (2) and (3) represent the merged continuum flux and error for the photometric V-band (denoted as P) and spectroscopic f$_{5100}$(denoted as S), measured in units of 10$^{\mathrm{-15}}$ erg cm$^{-2}$ s$^{-1}$ \AA$^{-1}$; (4) and (5) are the H$\alpha$ line flux and its error, also in units of 10$^{\mathrm{-15}}$ erg cm$^{-2}$ s$^{-1}$; (6) specifies the type of continuum flux, where P indicates photometric V-band flux and S denotes spectroscopic flux measured at 5100\,\AA.  The complete table is available in a machine-readable format, with a portion provided here for reference.
\end{tablenotes}
\end{table*}

\section{Analysis and Results}
\label{sec:Results}
\subsection{Variability}\label{sec:Variability}
We find the fractional variability amplitude ($F_{\mathrm{var}}$) for merging photometric V-band and spectroscopic optical continuum flux at 5100{\AA} ($f_{\mathrm{5100}}$) using the following equation \citep{RodriguezPascual1997}.
\begin{align}
F_{\mathrm{var}}= \dfrac{\sqrt{(\sigma^{2}-<\sigma^{2}_{\mathrm{err}}>)}}{<f>}
\end{align}

where $\sigma^{2}$ is the variance, $\sigma^{\mathrm{2}}_{\mathrm{err}}$ is the mean square error, and $<f>$ is the arithmetic mean of the light curves. The $F_{\mathrm{var}}$ for the merged continuum of 10$^{\mathrm{th}}$ March is 2.6 $\%$ and 11$^{\mathrm{th}}$ March is 7$\%$. This for the entire light curve, including both days, is 6.0$\%$. The maximum variability for H$\alpha$ emission line flux is 6.3 $\%$ for two days.

Additionally, the ratio of maximum to minimum flux variation ($R_{\mathrm{max}}$) is calculated for these light curves. Table \ref{tab:Variability_table} shows the variability statistics including median flux, $F_{\mathrm{var}}$ and $R_{\mathrm{max}}$ for 10$^{\mathrm{th}}$ March, 11$^{\mathrm{th}}$ March, and the entire light curve.
The variability and $R_{\mathrm{max}}$ are similar to the range found in \citet{Cho_2020}. The $R_{\mathrm{max}}$ ranges from 1.18 to 1.33 for the H$\alpha$ emission line and from 1.12 to 1.30 for the merged continuum light curve.
\begin{table}
  \caption{Light curve Variability statistics}
  \hspace{-1.5cm}
  \scalebox{0.8}{
  \begin{tabular}{@{}llcccc@{}}
    \toprule
    DAY & Light curve & Median & $R_{\mathrm{max}}$ & $F_{\mathrm{var}}$ ($\%$) \\
    (1) & (2) & (3) & (4) & (5)\\
    \midrule
    \multirow{2}{*}{10$^{\mathrm{th}}$ March}
    & F$_{\mathrm{5100}}$+V-band & $28.97 \pm 0.12$ & $1.12 \pm 0.01$ & $2.60 \pm 0.24$ \\
    & H$\alpha$ & $331.20 \pm 2.80$ & $1.18 \pm 0.04$ & $3.67 \pm 0.54$ \\
    \multirow{2}{*}{11$^{\mathrm{th}}$ March}
    & F$_{\mathrm{5100}}$+V-band & $26.90 \pm 0.27$ & $1.28 \pm 0.02$ & $6.89 \pm 0.58$ \\
    & H$\alpha$ & $299.73 \pm 1.85$ & $1.15 \pm 0.03$ & $2.70 \pm 0.43$ \\
    \multirow{2}{*}{10$^{\mathrm{th}}$+11$^{\mathrm{th}}$ March} 
    & F$_{\mathrm{5100}}$+V-band & $28.56 \pm 0.19$ & $1.30 \pm 0.02$ & $6.38 \pm 0.40$ \\
    & H$\alpha$ & $314.45 \pm 2.81$ & $1.33 \pm 0.04$ & $6.31 \pm 0.53$ \\
    \bottomrule
  \end{tabular}}
\label{tab:Variability_table}
\begin{tablenotes}
 \item \textbf{Note.} Columns are (1) Day; (2) Nature of the light curve; (3)  Median flux of the light curve in units of 10$^{-15}$ erg s$^{-1}$cm$^{-2}$ \AA$^{-1}$ for $f_{\mathrm{5100}}$, and 10$^{-15}$ erg s$^{-1}$cm$^{-2}$ for emission lines; and (4) the ratio of maximum to minimum flux variation; (5) fractional rms variability in percentage.
\end{tablenotes}
\end{table}

\subsection{Time lag analysis}
Technical issues resulted in interruptions in continuous observations, leading to gaps in the spectroscopic light curve on both nights, as evident in Fig. \ref{fig:lightcurve_V}. Consequently, the light curve is divided into four segments (two for each day) before being used to measure the lag. Specifically, for 10$^{\mathrm{th}}$ March, only the second part of the light curve is used, as the first part lacked sufficient data points and monitoring duration.

\subsubsection{ICCF and \small{JAVELIN}}
\label{sec:timelag analysis}
The lag between the continuum and H$\alpha$ emission line is measured using two methods, namely the Interpolated Cross-correlation function (ICCF\footnote{\url{https://bitbucket.org/cgrier/python_ccf_code/src/master/}}) \citep{Peterson_1998} and \small{JAVELIN}\footnote{\url{https://github.com/nye17/JAVELIN}} \citep{Zu2011, Zu2013}. These methods are widely used \citep[see;][and references therein]{Peterson_1998, Peterson_2004, Bentz2014ApJ...796....8B, Barth2015, Woo2024} with consistent results \citep{Edelson2019}. In ICCF, the nature of the correlation between the two datasets is identified through cross-correlation. The lag between the optical continuum and H$\alpha$ emission line flux is determined with a lag limit of -60 to 160 minutes for segments and -60 to 300 minutes for entire light curves. In ICCF, one light curve is linearly interpolated while keeping the other light curve fixed, and vice versa. Subsequently, the average of the two Cross-Correlation Functions (CCFs) yields the final ICCF.   
 
The Monte Carlo method of flux randomization (FR) and random subset selection (RSS) is used to measure the uncertainty in the lag \citep{Peterson_1998, Peterson_2004} with a median of the centroid ($\tau_{\mathrm{cent}}$; covering 0.8$\times$ r$_{\mathrm{max}}$; where r$_{\mathrm{max}}$ is the maximum correlation coefficient achieved) distribution as the lag. Table \ref{tab:lag-values} depicts the lag results with different methods of lag estimation.

\small{JAVELIN} is developed by \citet{Zu2011, Zu2013} to model the driving light curve with a DRW model. This DRW-modeled light curve utilizes the transfer function to derive the other related light curves, typically employing a top-hat function as the transfer function. Error estimation employs the Markov Chain Monte Carlo (MCMC) methodology to compute statistical confidence limits for each best-fit parameter. This approach simultaneously enables continuum and emission-line light curves modeling. The DRW process \citep{Kelly2009, Kelly2014} accurately predicts the nature of continuum variations in AGN for long, short, and even multiband light curves with some tenable deviations as seen \citep[e.g.,][]{Mchardy2006, Mushotzky2011}. Smaller uncertainties in lag estimation are provided by \small{JAVELIN} compared to the ICCF method \citep[see;][]{Grier2017, Edelson2019, Yu2020}. However, they might be more affected by seasonal gaps that could infer a negative false rate \citep[for more detail;][]{Woo2024}.

The results of the time series analysis are plotted in Fig. \ref{fig:lag plots}, where the upper left panel displays a segment of the continuum light curve, while the lower left panel shows a segment of the emission line light curve. The lag is measured for various segments of the emission line light curve concerning the entire continuum light curve for the respective days. Lag results were considered reliable when the ICCF correlation coefficient ($r_{\mathrm{max}}$) was more than 0.5 (as shown in Fig. \ref{fig:lag plots}). For a lag limit of -60 to 160 minutes, a lag of approximately 86 minutes is obtained for the second part of the 10$^{\mathrm{th}}$ March with \small{JAVELIN}, consistent with the ICCF method, which provided a lag of 85 minutes. The entire light curve of the same day peaked at 268 minutes with a lag limit of -60 to 300 minutes. The more considerable uncertainty in these cases is due to detecting two peaks in the lag distribution. On the following day, 11$^{\mathrm{th}}$ March, with the same lag limit of -60 to 160 minutes, the lag for the first segment was 93 and 103 minutes with \small{JAVELIN} and ICCF, respectively. 

For the second segment of 11$^{\mathrm{th}}$ March, the results from ICCF is 124 minutes. However, \small{JAVELIN} evidences two peaks in the lag distribution, which are around 0 and 122 minutes. The more substantial first peak leads to a lag of $\sim$ 3 minutes with considerable uncertainty. However, if we remove the first peak around 0 minutes, it is evident that the lag result for 0 to 160 minutes matches with the ICCF lag stated in Table \ref{tab:lag-values}. The entire light curve for 11$^{\mathrm{th}}$ March shows slightly different results using both methods with a lag limit of -60 to 300 minutes to be 185 minutes with \small{JAVELIN} and 150 minutes with ICCF. The lag is also calculated by combining both days' light curves, shown in the Table \ref{tab:lag-values} for the limit -60 to 300 minutes. Lower cross-correlation coefficient $r_{\mathrm{max}}$ is encountered due to noise and gaps in the light curve. The results for a few chunks of the light curve are obtained with detrending, discussed in Section \ref{sec:detrended}.

\subsubsection{Detrending the light curve}
\label{sec:detrended}
Understanding the short-term intrinsic behavior is crucial for understanding the variability and estimating the reliable lag. However, long-term trends, resulting, for example, due to red-noise leakage \citep{Welsh1999}, pose a challenge when analyzing short-term behavior and can introduce bias into cross-correlation lag results. An effective approach is to eliminate these long-term trends, which not only aids in understanding short-term variability but also improves lag estimation \citep[see,][]{Zhang2019,Woo2024}. To achieve this, we fit a straight line to each light curve and subtract it from the original, removing the trend. Subsequently, we cross-correlate the continuum and line light curves to measure the lag between them. It's worth noting that higher-order polynomials can introduce unnatural or false signals and patterns into the light curve, potentially impacting lag estimates.

\begin{figure*}
\begin{minipage}[t]{0.5\textwidth}
\includegraphics[scale=1,width=3.5in,height=2in]{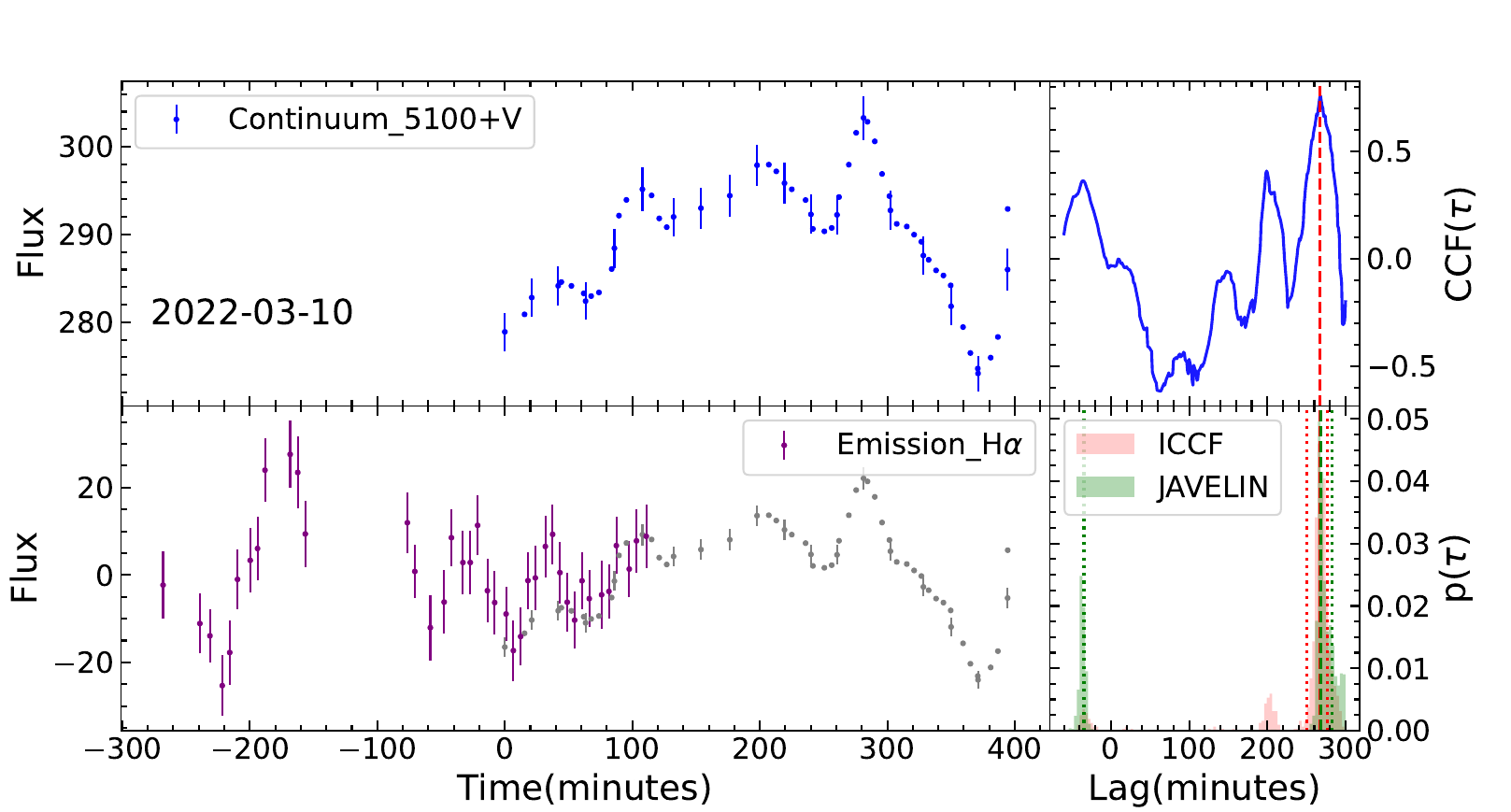}
\end{minipage}%
\begin{minipage}[t]{0.5\textwidth}
\includegraphics[scale=1,width=3.5in,height=2in]{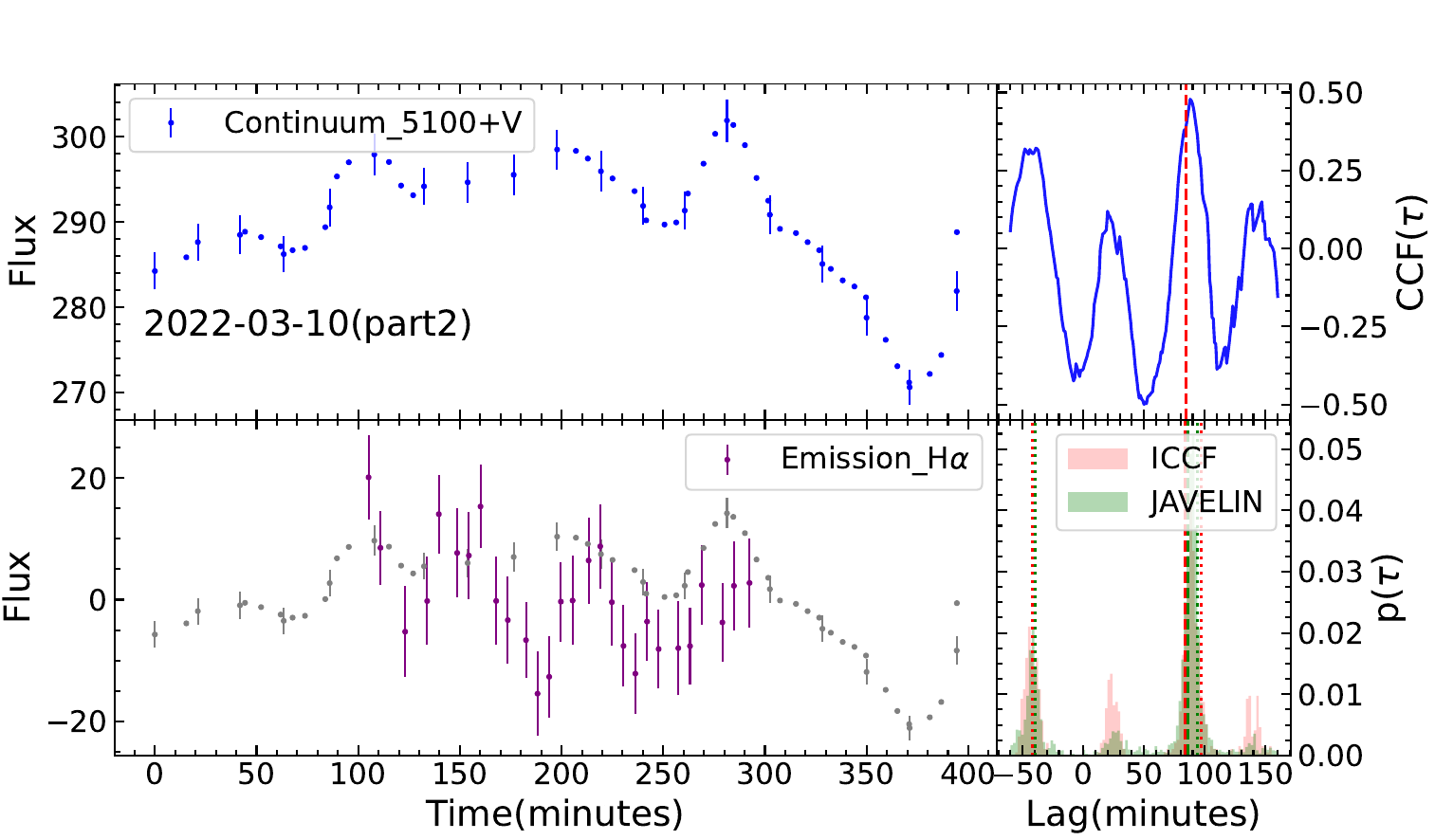}
\end{minipage}
\begin{minipage}[t]{0.5\textwidth}
\includegraphics[scale=1,width=3.5in,height=2in]{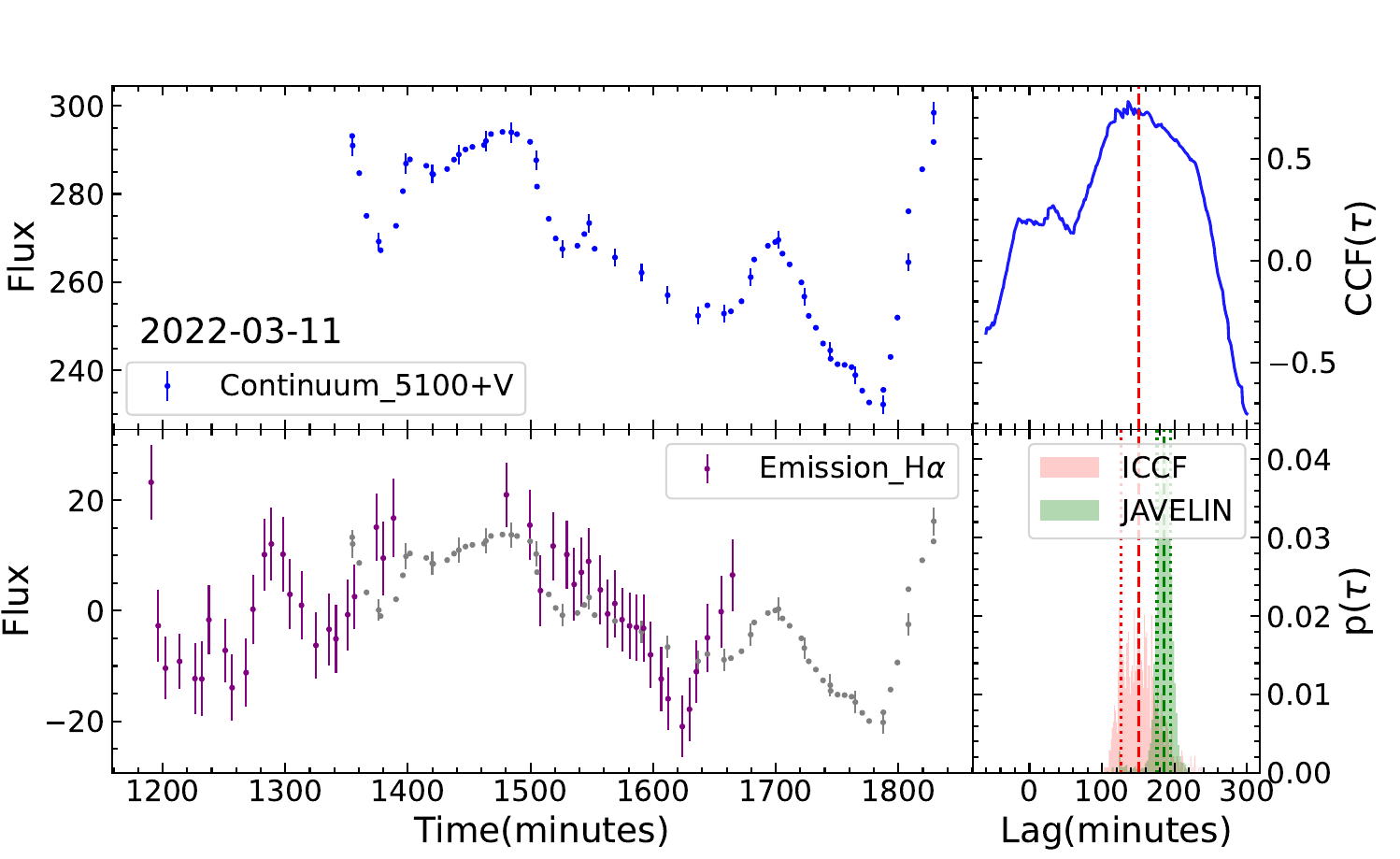}
\end{minipage}%
\begin{minipage}[t]{0.5\textwidth}
\includegraphics[scale=1,width=3.5in,height=2in]{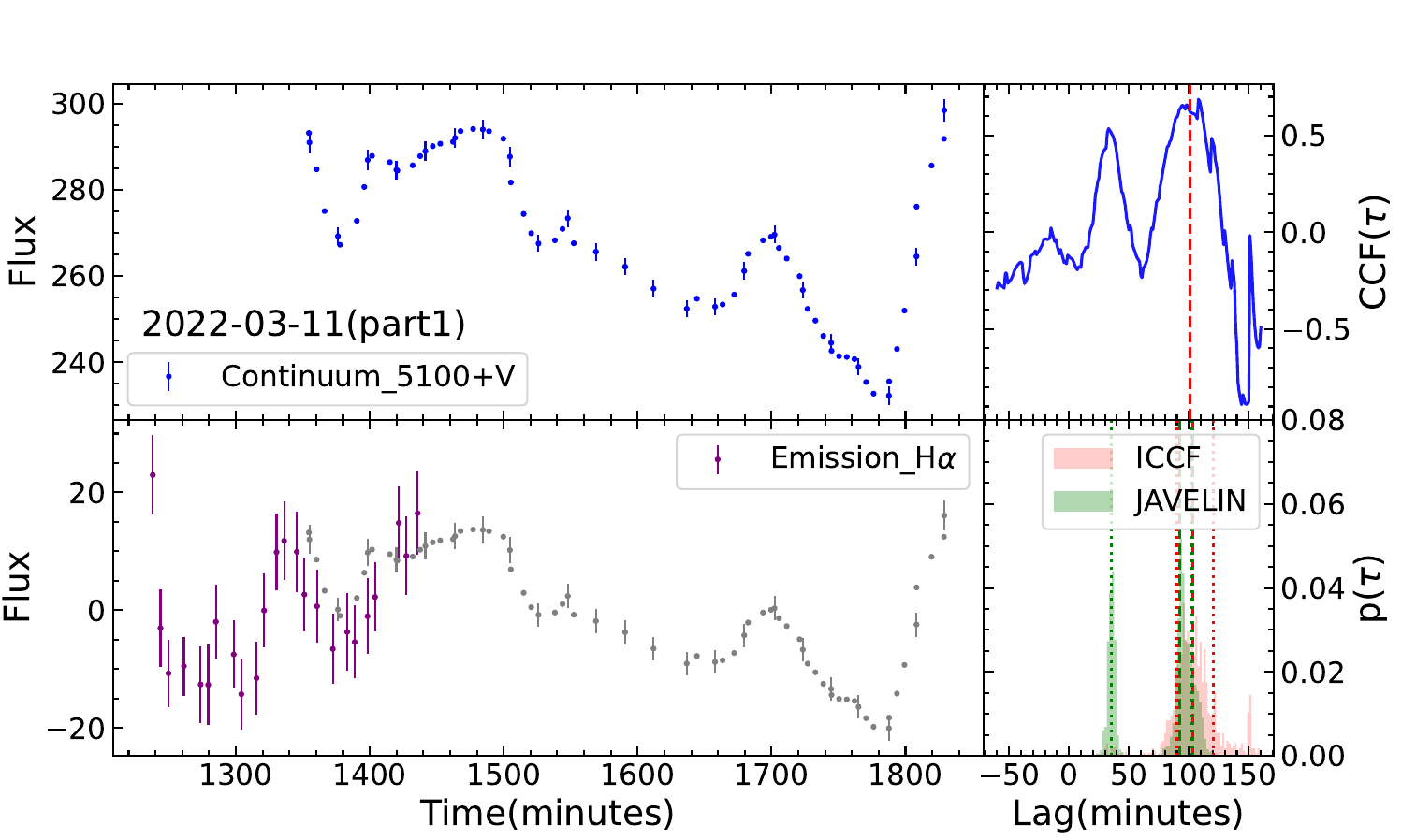}
\end{minipage}
\begin{minipage}[t]{0.5\textwidth}
\includegraphics[scale=1,width=3.5in,height=2in]{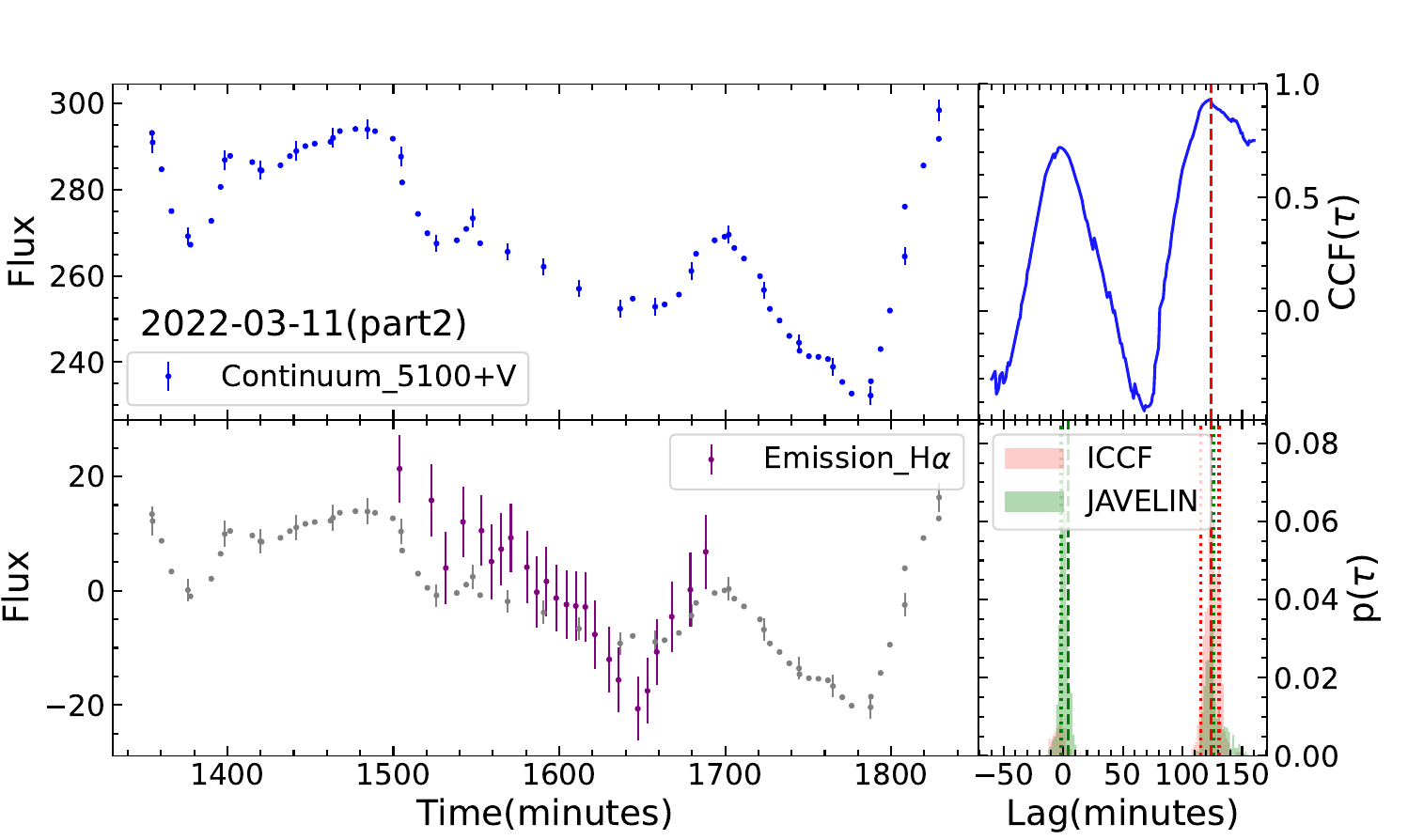}
\end{minipage}
\begin{minipage}[t]{0.5\textwidth}
\includegraphics[scale=1,width=3.5in,height=2in]{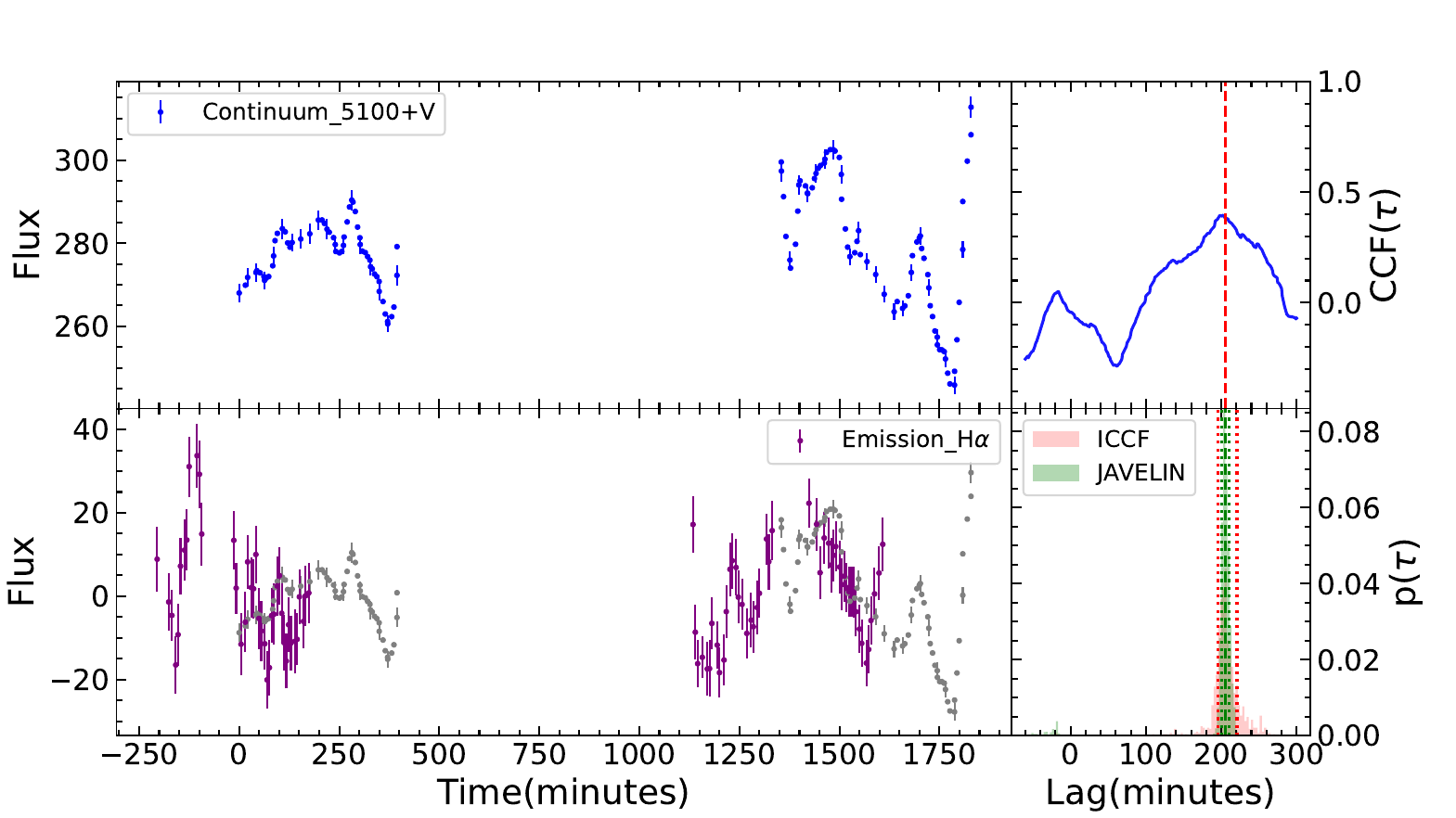}
\end{minipage}%
\caption{The lag results are plotted; the upper left panel is the merged V-band and continuum flux at 5100{\AA} in the unit of 10$^{\mathrm{-17}}$ erg cm$^{\mathrm{-2}}$s$^{\mathrm{-1}}$\AA$^{\mathrm{-1}}$, and the lower left panel showing the emission line light curve in the unit of 10$^{\mathrm{-17}}$ erg cm$^{\mathrm{-2}}$s$^{\mathrm{-1}}$. The mean-subtracted light curves shown are matched by normalizing the continuum light curve and shifting the emission line light curve to the final adopted lag values mentioned in Table \ref{tab:lag-values}.} The upper right depicts the ICCF coefficient (r) resulting from the continuum and emission line light curve with respect to lag in minutes. The lower right panel are the lag histograms obtained using methods ICCF and \small{JAVELIN}. Here, the subplots 1, 2, and 6 show the detrended light curves and their respective detrended lag results.
\label{fig:lag plots}
\end{figure*}

In some parts of the light curves, two prominent peaks conflicted with the lag estimation through ICCF and \small{JAVELIN}. Consequently, we detrended the 10$^{\mathrm{th}}$ March light curve (entire and in parts), which improved the cross-correlation coefficient ($r_{\mathrm{max}}$). In Fig. \ref{fig:lag plots}, the detrended light curves are shown for the 10$^{\mathrm{th}}$ March second part and the entire light curves of the continuum and H$\alpha$ emission line. Similarly, the lag calculation for the entire two days is significantly affected by detrending and results in one peak of around 205 minutes shown in section \ref{sec:appendix}. However, for 11$^{\mathrm{th}}$ March part 2, the detrending does not pose any difference in the results, especially with \small{JAVELIN}.
\begin{table*}
\caption{Lag Table}
\hspace{-2cm}
\begin{tabular}{c@{\hspace{5mm}}c@{\hspace{5mm}}c@{\hspace{5mm}}c@{\hspace{5mm}}c@{\hspace{5mm}}c@{\hspace{5mm}}c@{\hspace{5mm}}c@{\hspace{5mm}}c@{\hspace{5mm}}c@{}}
\hline \hline
\multicolumn{2}{c}{\textbf{DAY}} && \multicolumn{2}{c}{\textbf{ICCF}} && \textbf{\small{JAVELIN}} && \multicolumn{2}{c}{\textbf{PyI$^{\mathrm{2}}$CCF}}\\ 
\cmidrule{1-2} \cmidrule{4-5} \cmidrule{7-7} \cmidrule{9-10}
Date& Segment& &Lag&r$_{\mathrm{max}}$&&Lag&&Lag&p value\\
(1)&(2)&&(3)&(4)&&(5)&&(6)&(7)\\ \hline
\multirow{2}{*}{10$^{\mathrm{th}}$ March} & Entire$^{\star}$ && $267.3^{+9.2}_{-16.7}$& 0.70&& $268.4^{+14.6}_{-302.8}$ && $268.0^{+8.0}_{-10.2}$&0.31 \\
& Part 2$^{\star}$ && $84.5^{+11.4}_{-126.6}$&0.45&&  $86.1^{+7.6}_{-125.3}$&& $86.5^{+7.5}_{-130.0}$&0.60\\
\multirow{2}{*}{11$^{\mathrm{th}}$ March} & Entire && $150.5^{+25.8}_{-23.7}$&0.78 && $185.3^{+9.7}_{-9.3}$ && $148.5^{+26.4}_{-24.6}$&0.35\\
& Part 1 && $103.1^{+17.4}_{-13.3}$&0.60&& $92.6^{+10.3}_{-57.4}$ && $101.1^{+15.4}_{-12.5}$&0.87\\
& Part 2 && $124.2^{+6.6}_{-8.6}$&0.90&&  [$122.3^{+9}_{-4}$]$^{\dagger}$&&$125.0^{+6.2}_{-6.1}$&0.04 \\
10$^{\mathrm{th}}$ March+11$^{\mathrm{th}}$ March& Entire$^{\star}$ &&205.6$^{+16.6}_{-11.1}$ &0.38 &&205.0$^{+5.5}_{-4.4}$ &&205.0$^{+17.0}_{-9.3}$ &0.54 \\
\hline
\end{tabular}
\label{tab:lag-values}
\begin{tablenotes}
\item $^{\star}$ Detrended light curves.
\item $^{\dagger}$ Considered second peak for lag estimation.
\item \textbf{Note.} Lag in rest frame between combined continuum (V-band and $f_{5100}$) vs. H$\alpha$ light curves obtained using ICCF, \small{JAVELIN}, and PyI$^{\mathrm{2}}$CCF for different days and segments. Columns: (1) Date of observation; (2) Segment of the lightcurve; (3) ICCF lag in minutes; (4) Cross-Correlation coefficient r$_{\mathrm{max}}$; (5) \small{JAVELIN} lag in minutes; (6) PyI$^{\mathrm{2}}$CCF lag in minutes; (7) PyI$^{\mathrm{2}}$CCF Null hypothesis value (p). 
\end{tablenotes}
\end{table*}

\subsubsection{Simulations}
\label{sec:simulations}
The publicly available PyI$^{\mathrm{2}}$CCF code\footnote{\url{https://github.com/legolason/PyIICCF/}} developed by \citet{Guo2022} is used, implementing the method described in \citet{U2022}. This method assesses the significance of lag measurements or verifies the reliability of the ICCF method. It is grounded on the null hypothesis, which posits that when two random uncorrelated light curves are cross-correlated, the resulting $r_{\mathrm{max}}$ should be greater than or equal to the observed cross-correlation ($r_{\mathrm{max},\mathrm{obs}}$) obtained from real light curves. Mock light curves are generated by the code based on the DRW model with the same noise and cadence as the actual light curves \citep[see also,][]{U2022,pandey_spectroscopic_2022, Buitrago2023, Woo2024}. Specifically, 1000 mock light curves are generated for the continuum and H$\alpha$ emission line light curves. The lag values in minutes alongside the null hypothesis value ($p$) are presented in the last column of Table \ref{tab:lag-values}. Certain light curve parts in Table \ref{tab:lag-values} exhibit more noise and, consequently, poorer quality, as reflected in the higher p-value. Applying the criteria for lag reliability $p \leq 0.2$ \citep{U2022, Guo2022, Woo2024} and $r_{\mathrm{max}} > 0.5$, we identified the second segment of 11$^{\mathrm{th}}$ March to provide the most reliable lag, which is $125.0^{+6.2}_{-6.1}$ minutes and used it for the black hole mass measurement. We caution that the above may not be the best criteria for lag reliability.

\subsection{Line width and black hole mass measurement}
\label{sec:BHmass}
To construct the mean spectrum, we first aligned all spectra to match the resolution of the H$\alpha$ emission line using mapspec\footnote{\url{https://github.com/mmfausnaugh/mapspec}} \citep{MAPSPEC}. Consequently, the mean spectrum is constructed for each part of the observations using the following approach. 
\begin{align}
\bar{F(\lambda)} =\dfrac{1}{N}\sum_{i=0}^{N-1}F_{i}(\lambda),
\end{align} 
where $F_{\mathrm{i}}(\lambda)$ is the $i$-th spectrum of the $N$ spectra that comprise the database. 
For 10$^{\mathrm{th}}$ March, we acquired 40 spectra, with 28 available in the second segment. Similarly, for 11$^{\mathrm{th}}$ March, 24 spectra were obtained for each segment, resulting in 48 spectra for the entire night. Fig. \ref{fig:Mean-Spectrum} displays one such mean spectrum obtained from 11$^{\mathrm{th}}$ March.

As mentioned in section \ref{sec:decomp}, the mean spectrum is also decomposed into broad and narrow components of the H$\alpha$ line. Then the best-fit broad H$\alpha$ component is used to measure the full width at half maximum (FWHM) and $\sigma_{\mathrm{line}}$ in the wavelength region of 6500-6600{\AA}.

Following \citet{Peterson_2004}, the FWHM is calculated by finding half maximum both from the left ($\lambda_{\mathrm{left}}$) and right side ($\lambda_{\mathrm{right}}$) of the curve. Eventually, the FWHM is the $\lambda_{\mathrm{right}}$- $\lambda_{\mathrm{left}}$ wavelengths, whereas, for calculation of $\sigma_{\mathrm{line}}$, the flux weighted line center is first determined as follows:
\begin{align}
\lambda_{\mathrm{0}} =\dfrac{\int \lambda f_{\mathrm{\lambda}}\,d\lambda}{\int f_{\mathrm{\lambda}}\,d\lambda}
\end{align}
and the line dispersion as follows:
\begin{align}
\sigma_{\mathrm{line}}^{2} =\dfrac{\int \lambda^{2} f_{\mathrm{\lambda}}\,d\lambda}{\int f_{\mathrm{\lambda}}\,d\lambda} - \lambda_{\mathrm{0}}^{2}
\end{align}

The FWHM and $\sigma_{\mathrm{line}}$ are measured using the Monte Carlo bootstrap technique \citep{Peterson_2004}, where 5000 realizations are run to measure them after the $N$ spectra were chosen from the pool of $N$ spectra. The endpoints of the H$\alpha$ region (6500-6600{\AA}) are randomly adjusted within a range of $\pm$10{\AA} from the initially selected values. The distribution for FWHM and $\sigma_{\mathrm{line}}$ is obtained after running 5000 realizations, estimating the uncertainties in the line width and the black hole mass measurement. The FWHM and $\sigma_{\mathrm{line}}$, along with their measured uncertainties (the 16$^{\mathrm{th}}$ and the 84$^{\mathrm{th}}$ percentiles of the distribution) are presented. Finally, the FWHM and $\sigma_{\mathrm{line}}$ are corrected with the instrumental resolution.

Given that the gravitational potential of the black hole influences the movement of gas within the Broad Line Region (BLR), the mass of the black hole can be measured by multiplying the size of the BLR ($R_{\mathrm{BLR}}$) and the velocity width of the broad emission lines ($\Delta$V), using the virial relation:

\begin{align}\label{eq:virial eqn}
M_{\mathrm{BH}} =  \dfrac{f \times R_{\mathrm{BLR}} (\Delta V)^{2}}{G}
\end{align}

where $f$ is a dimensionless scale factor that depends on the kinematics and geometry of BLR gas clouds. The BLR size is directly measured with lag($\tau$) in minutes using $R_{\mathrm{BLR}}$=c$\tau$ from Table \ref{tab:lag-values} with c as the speed of light.

The scale factor ($f$) is taken from \citet{Woo2015}, which is 1.12 and 4.47 when the line width is chosen as FWHM and $\sigma_{\mathrm{line}}$, respectively. The resolution corrected $\sigma_{\mathrm{line}}$ and FWHM from the mean spectrum are measured to be 544.7$^{+22.4}_{-25.1}$ km s$^{-1}$ and 810.2$^{+86.8}_{-91.8}$ km s$^{-1}$, respectively. Combining the PyI$^2$CCF lag of 125 minutes (11th March, Part2) with the $\sigma_{\mathrm{line}}$ and FWHM, the black hole mass is calculated to be 2.2$^{+0.2}_{-0.2} \times 10^{4} M_{\mathrm{\odot}}$ and 1.2$^{+0.3}_{-0.2} \times 10^{4} M_{\mathrm{\odot}}$, respectively. 

Since $\sigma_{\mathrm{line}}$ is less sensitive to the peak and provides more accurate black hole mass measurement than FWHM \citep[e.g.,][]{Peterson_2004,2014SSRv..183..253P}, a black hole mass of $2.2^{+0.2}_{-0.2}\times 10^{4}M{\mathrm{\odot}}$ is found for NGC 4395 as the best measurement.

\section{Discussion}
\label{sec:Discussion}
\subsection{Previous Studies}
An extensive range of black hole mass for NGC 4395 has been reported in the literature, e.g., the mass of the black hole was calculated to be 1.2 $\times$ 10$^{5}$ $M_{\mathrm{\odot}}$ by \citet{kraemer1999}, using photoionization modeling with an FWHM of H$\beta$ as 1500 km s$^{-1}$, while \citet{Filipenko2003} provided a mass range $250 M_{\mathrm{\odot}} \le M_{BH} \le 6 \times 10^{6} M_{\mathrm{\odot}}$ based on the Eddington limit and dynamics of the stellar cluster. Utilizing X-ray, UV, and optical photometric data points along with spectroscopic observations, \citet{Desroches2006} confirmed the reprocessing model and measured a black hole mass of 3 $\times$ 10$^{5}$ $M_{\mathrm{\odot}}$ using HeII, H$\alpha$, H$\beta$, and H$\gamma$ emission lines. The gas dynamics of the central region for NGC 4395 have been studied by a few authors, such as \citet{brok_measuring_2015} and \citet{Brum2019}, to estimate the $\sigma_{\mathrm{\star}}$ that can help constrain the black hole mass. Unfortunately, the disk continuum emission might have been dominated by the CO absorption bandhead, which makes it insignificant and prevents the measurement of the stellar dispersion velocity. Although with single epoch spectroscopic data obtained using Gemini GMOS-IFU, the black hole mass of NGC 4395 was found as 2.5 $\times$ 10$^{5}M_{\mathrm{\odot}}$ with H$\alpha$ FWHM as 785 km s$^{-1}$ using nuclear spectrum \citep{Brum2019}. Whereas, Gemini NIFS spectrum was used by \citet{brok_measuring_2015} to measure the best-fit black hole mass of $4^{+8}_{-3}\times 10^{5} M_{\mathrm{\odot}}$(3$\sigma$ uncertainties) with combined modeling of the stellar populations of the nuclear star cluster (NSC) and the dynamics of the molecular gas.

Several attempts of RM have been made for NGC 4395. For example, \citet{peterson_multiwavelength_2005} performed RM using the CIV $\lambda$1549 emission line, providing a BLR size of 48-66 light minutes. Combining the lag with FWHM ($\sim$ $\sigma_{\mathrm{line}}$) of $\sim$ 3000 km s$^{-1}$ from the rms spectra, they measured a black hole mass of 3.6$\pm$1.1 $\times$ 10$^{5}$ $M_{\mathrm{\odot}}$. It's important to mention that the CIV line depicted in \citet{peterson_multiwavelength_2005} exhibits an irregular line profile, coupled with a noisy root mean square (rms) spectrum, which could potentially hinder its accurate measurement \citep[as discussed in][]{cho2021}. The first reliable black hole mass was reported by \citet{woo_10000-solar-mass_2019}, who determined $\sim$80 min lag based on the narrow-band photometry RM results and measured the $\sigma_{\mathrm{line}}$ of H$\alpha$ as 426$\pm$2.5 km s$^{-1}$ from the mean spectrum, obtaining black hole mass as $\sim$10,000 M$_{\mathrm{\odot}}$. However, as mentioned in the section \ref{sec:intro}, the continuum contribution in the narrow-band H$\alpha$ filter can be a dominant source of uncertainty in the lag measurement. \citet{woo_10000-solar-mass_2019} assumed the variability of continuum in the H$\alpha$ band (6450-6650{\AA}) is similar to what they measured in the photometric continuum V-band (4800-6500{\AA}). Further, they convolved the transmission curve of the MDM H$\alpha$ filter with the mean GMOS spectrum and calculated a mean continuum contribution of 18.3 \%. The continuum fraction in each H$\alpha$ epoch was estimated by scaling the V-band variability by a factor of 0.183, subtracted from each H$\alpha$ epoch. It is worth noting that this represents the maximum calculated continuum contribution, as the continuum variability in the H$\alpha$ filter is expected to be no greater than that in the V-band filter. \citet{Cho_2020} varied the scaling parameter by assuming the continuum inside the H$\alpha$ filter has a similar variability behavior to the photometric continuum V-band. The variability is varied with a range of 0-100 \% to that of the H$\alpha$ emission line flux using the variability amplitude parameter equation presented in their section 3.3.1. However, their PRM analysis is affected by baseline mismatches between H$\alpha$ and V-band observations, weak variability, and significant uncertainty in H$\alpha$ photometry, all impacting lag measurement. By decomposing and separating the emission lines with the continuum using flux-calibrated spectra, the spectroscopic RM can accurately predict the variability and contribution of the continuum and further the lag measurement.

This black hole mass was updated as 1.7 $\pm$ 0.3 $\times$ 10$^{4}$ $M_{\mathrm{\odot}}$ by \citet{cho2021} as the $\sigma_{\mathrm{line}}$ of H$\alpha$ was re-measured as 586$\pm$19 km s$^{-1}$ based on a higher quality spectrum by fitting the broad component with a double Gaussian. \citet{cho2021} attempted spectroscopic RM of NGC 4395. However, they could not obtain a reliable lag measurement due to bad weather losses and insignificant detection of variability structure in the light curve. At the same time, they constrained the upper limit of the reverberation black hole mass as 4 $\times$ 10$^{4}$ $M_{\mathrm{\odot}}$ with lag as less than 3 hours. Recently, \citet{Gu2024} employed broad-band PRM by analyzing the g- and r-band data of NGC 4395 from recent studies \citep{Montano2022, McHardy2023} measuring H$\alpha$ lag of 40 to 90 min. We performed spectroscopic reverberation mapping and performed detailed spectral decomposition to construct the emission line light curve. These good-quality light curves allowed us to successfully estimate the BLR size of 125 light minutes using both ICCF and JAVELIN. Moreover, we measured velocity dispersion and constrained the black hole mass to be $2.2^{+0.2}_{-0.2}\times 10^{4}M_{\mathrm{\odot}}$ which is consistent with the findings of \citet{woo2019}, \citet{Cho_2020} and \citet{cho2021}.

The FWHM for H$\alpha$ was used as 1500$\pm$500 km s$^{-1}$ (lag $\sim$ 3.6 hours) by \citet{Edri2012}, resulting in a black hole mass estimation, i.e., 4.9 $\pm$ 2.6 $\times$ 10$^{4}$ $M_{\mathrm{\odot}}$. They further argued that FWHM provides a better measurement for line width than taking the second moment, i.e., $\sigma_{\mathrm{line}}$, because the broad lines become narrower, as the single Gaussian fit might not suit it. Instead, modeling the broad component with a Lorentzian might be more appropriate \citep{Kollatschny2011}. However, \citet{cho2021} and \citet{woo2019} suggested that H$\alpha$ fits very well with a double Gaussian model. It should be noted that this study used a double Gaussian to model the broad H$\alpha$ line, similar to modeling with a Lorentzian profile.

The RMS spectrum of NGC 4395 was used by \citet{peterson_multiwavelength_2005} and \citet{Desroches2006} to measure line width and calculate the black hole mass. On the other hand, it was noted by \citet{woo2019} that the $\sigma_{\mathrm{line}}$ is approximately similar when using both RMS and mean spectra. In our study, the RMS spectrum was noisy, and consequently, FWHM and $\sigma_{\mathrm{line}}$ were obtained from the mean spectrum. Our results are slightly lower than previous estimates, with FWHM measured at 810.2$^{+86.8}_{-91.8}$ km s$^{-1}$ and $\sigma_{\mathrm{line}}$ at $544.7^{+22.4}_{-25.1}$ km s$^{-1}$ for the second part of the observations on 11$^{\mathrm{th}}$ March. Furthermore, our H$\alpha$ line width, denoted as $\sigma_{\mathrm{line}}$, is 7\% narrower compared to the H$\alpha$ line profile presented by \citet{cho2021}, which was modeled using two broad components. Conversely, it is 28\% wider than the $\sigma_{\mathrm{line}}$ value of 426 km s$^{-1}$ reported by \citet{woo_10000-solar-mass_2019}, derived from a single Gaussian model.

The choice of the scale factor ($f$) value can significantly affect the mass measurement. The $f$ value of 5.5 adopted by \citet{peterson_multiwavelength_2005} was determined empirically by \citet{Onken2004} with $\sigma_{\mathrm{line}}$ as line width estimator. This value of $f$ was derived under the assumption that the relationship between the central black hole mass and $\sigma_{\mathrm{\star}}$, the $M_{\mathrm{BH}}$–$\sigma_{\mathrm{\star}}$ relation, is the same for both quiescent and active galaxies. A much smaller value was chosen by \citet{Edri2012} as 0.75 with FWHM as linewidth, based on the assumption of simple Keplerian motion, resulting in a smaller measured black hole mass. Similar to \citet{Onken2004}, \citet{Woo2015} added a large number of sources having RM-based masses to $M_{\mathrm{BH}}$–$\sigma_{\mathrm{\star}}$ relation and found the $f$ to be 1.12 and 4.47 with FWHM and $\sigma_{\mathrm{line}}$ as line width values, respectively. Hence, we have used $f=$ 4.47 \citep{Woo2015} with $\sigma_{\mathrm{line}}$ to calculate black hole mass of NGC 4395. Dynamical modeling of reverberation mapping data could help to better measure the $f$ value for NGC 4395.

\subsection{Host-galaxy Contribution}
The host galaxy and a nuclear star cluster (NSC) contribute to the measured luminosity. The host galaxy contribution (including the study of NSC) has been separated from the AGN core by previous authors \citep{Filipenko2003, Cameron2012} using \small{GALFIT} software \citep{Peng2002, Peng2010} or employing different models \citep{Cho_2020}. The \small{GALFIT} software is used to fit the galaxy profile by modeling it. An 18 $^{\mathrm{\prime \prime}}$ by 15 $^{\mathrm{\prime \prime}}$ rectangle was used by \citet{Cameron2012} to fit the central engine of NGC 4395 with three profile model components: a nuclear point source, a Sersic component, and an exponential disc. The Sersic component models the NSC; the bright irregular features and Galactic stars were masked during fitting. The nuclear point source was subtracted, and the non-nuclear flux density in the $u$, $v$, and $b$ bands was estimated. Additionally, two profile model components were used by \citet{Cho_2020}, a point source and an exponential disk, estimating the host galaxy contribution to the total flux at 16$\%$. 

The mean spectrum of the second segment of 11$^{th}$ March light curve was used, and the monochromatic luminosity at 5100{\AA} i.e., $\lambda L_{\mathrm{\lambda}}$ measured as 2.87$\pm$0.01 $\times$ 10$^{40}$ erg s$^{-1}$.
The image decomposition using GALFIT on the mean V-band images was performed. The model components used are a point spread function to model the AGN core within 8 $^{\mathrm{\prime \prime}}$ and an exponential disk with addition to the Sersic profile for the host galaxy modeling. We found the host contribution of 22.6$\%$, slightly larger than the values reported by \citet{Cho_2020}. Considering the NSC flux of $\lambda L_{\mathrm{\lambda ;NSC}}$ = 3.59 $\times 10^{39}$ erg s$^{-1}$ from stellar cluster \citep{Carson_2015}, we measured the host and NSC subtracted final $\lambda L_{\mathrm{5100}}$ is 1.86$\pm$0.01 $\times$ 10$^{40}$ erg s$^{-1}$. 

\subsection{Radius-Luminosity relation}
\label{sec:R_L relation}
\begin{figure}
    \centering
    \includegraphics[height=8cm, width=8.5cm]{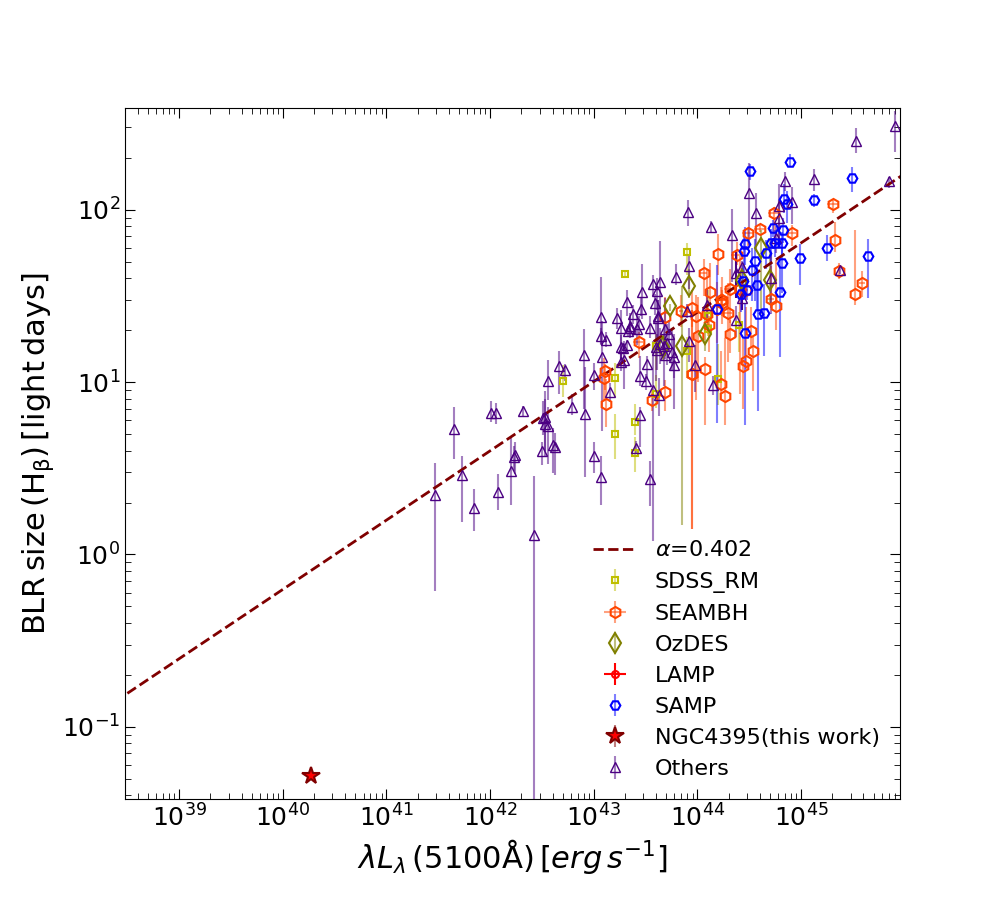}
    \caption{The plot depicts the H$\beta$ BLR size vs. optical luminosity at 5100{\AA}. NGC 4395 lies at a low luminosity end with H$\beta$ BLR size of 74.8 light minutes. The other RM sources, such as SEAMBHs \citep[][orange open circles]{Du_2016, Du_2018,Hu2021,Li2021}, SDSS-RM \citep[][yellow open squares]{Grier2017}, OzDES \citep[][green open diamonds]{Malik2023}, LAMP \citep[][red open circles]{U2022}, SAMP \citep[][blue open hexagons]{Woo2024}, and others \citep[including,][in purple open triangles]{bentz_low-luminosity_2013, Park_2017, Rakshit2019, Bonta2020, Rakshit2020A&A, pandey_spectroscopic_2022} are shown. The brown dashed line shows the best-fit R-L relation as obtained by \citet{Woo2024}.}
    \label{fig:lag-lum}
\end{figure}

To compare NGC 4395 with other RM sources, it has been plotted in the well-known H$\beta$ BLR size vs optical luminosity at 5100{\AA} established for AGNs. The measured H$\alpha$ lag ($\tau_{\mathrm{H\alpha}}$) has been converted to H$\beta$ lag ($\tau_{H\beta}$) lag using $\tau_{\mathrm{H\alpha}}:\tau_{H\beta}$=1.68:1 provided by \citet{Cho2023}. The measured H$\beta$ lag of 74.8 minutes has been plotted against $L_{\mathrm{5100}}$ of 1.86$\pm$0.01 $\times$ 10$^{40}$ erg s$^{-1}$ in Fig. \ref{fig:lag-lum}. Similar to \citet{Cho_2020}, NGC 4395 has been found to have a small offset from the fitted slope in the $R_{\mathrm{BLR}}$ - L relation. The reasons for such deviation could be partly due to either the uncertainty in the AGN luminosity measurement or intrinsically, the $R_{\mathrm{BLR}}$ - L relation for typical broad line AGNs is not valid for low-luminosity objects. Indeed, the measured luminosity is in the upper range compared to the previous works. For example, $L_{\mathrm{5100}}$ of 5.9 $\times$ 10$^{39}$ erg s$^{-1}$ was found by \citet{peterson_multiwavelength_2005}, while \citet{Cho_2020} calculated it to be 8.52 $\times 10^{39}$ erg s$^{-1}$, and 6.6 $\times 10^{39}$ erg s$^{-1}$ was measured by \citet{Filipenko2003}.

\begin{figure}
    \centering
    \includegraphics[height=8cm, width=8.5cm]{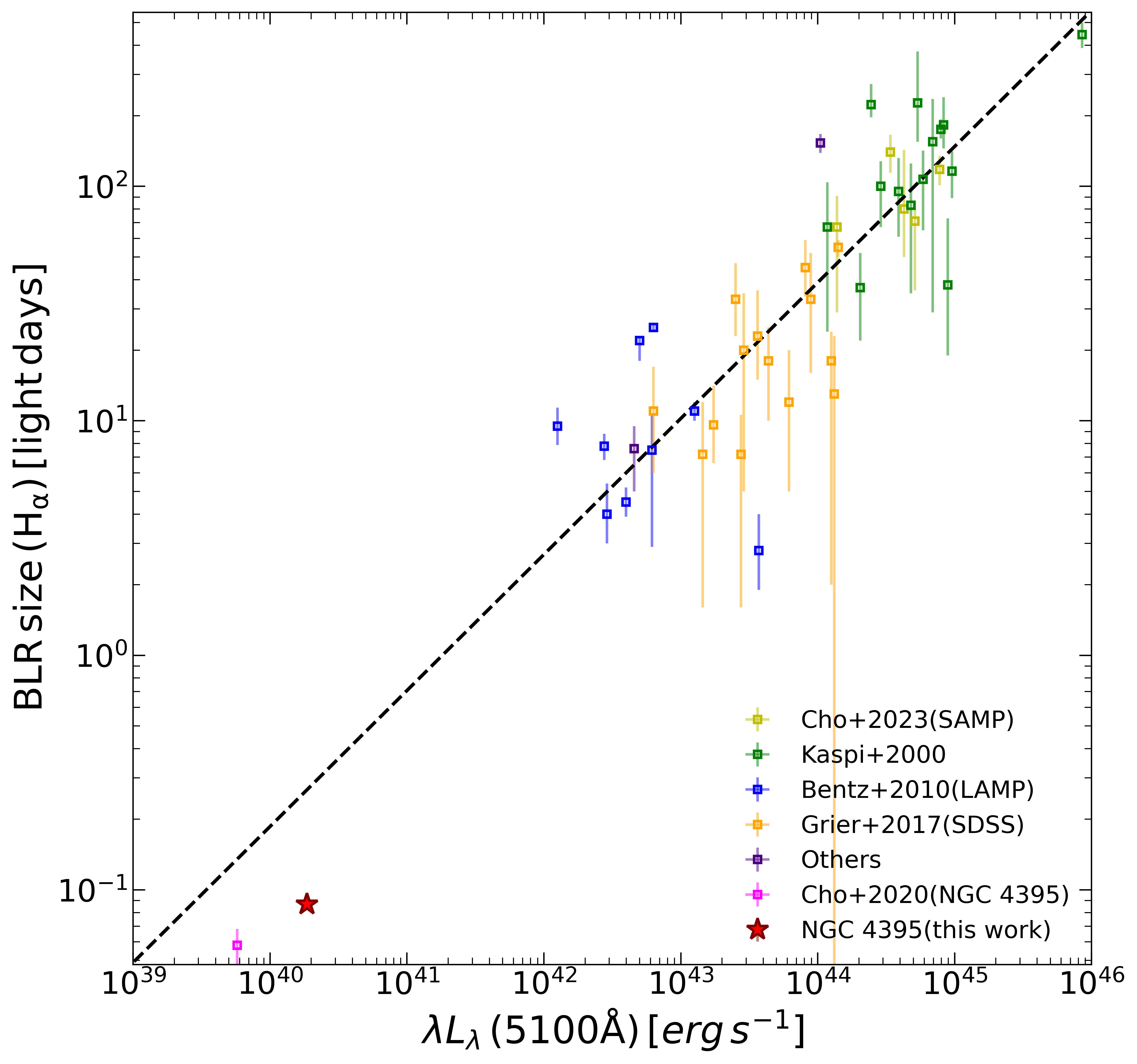}
    \caption{The H$\alpha$ BLR size vs. monochromatic continuum luminosity at 5100{\AA} is plotted. Various H$\alpha$ RM sources \citep{Kaspi2000, Bentz2010, Grier2017, Cho_2020, Woo2024} along with others \citep[such as,][]{Sergeev2017, Feng_2021, Li_2022} and the best-fit relation from \citet{Cho2023} are plotted. The source NGC 4395 is plotted with our best measurement of 125 light minutes in comparison to the 84 minutes lag obtained by \citet{Cho_2020}.}
    \label{fig:Ha_RL}
\end{figure}

The uncertainty could also arise from the choice of plotting the H$\alpha$ line lag, whereas the BLR size and luminosity relation are based on the H$\beta$ line lag. Recently, a best-fit size-luminosity relation, including NGC 4395, was obtained by \citet{Cho2023} as shown in Fig. \ref{fig:Ha_RL}. According to this, the expected H$\alpha$ lag for NGC 4395 is 374 minutes for our measured $L_{\mathrm{5100}}$, i.e., a factor of $\sim$ three times larger than our estimated lag of 125 minutes. With the Greene \& Ho + Bentz(GH+B) relation mentioned in \citet{Cho2023}, the $\tau_{\mathrm{H\alpha}}$ is estimated as 252 minutes with L$_{\mathrm{H\alpha}}$ as 6.64 $\times$ 10$^{37}$ erg s$^{-1}$. Both of these estimates are approximately similar and more than twice longer than our measurement for $\tau_{\mathrm{H\alpha}}$. This agrees with \citet{Cho2023} that the GH+B relation provides the lag results overestimated by twice for IMBHs or low-luminosity AGNs with L$_{\mathrm{H\alpha}}$ $<$ 10$^{42}$ erg s$^{-1}$. This can further overestimate black hole masses by 2-3 times for IMBHs. \citet{Zhang2007} noted that low-luminosity AGNs, characterized by L$_{\mathrm{H\alpha}}$ $<$ 10$^{41}$ erg s$^{-1}$, exhibit a lower ionization parameter and lower electron density in the BLR, potentially leading to deviations from the $R_{\mathrm{BLR}}-L$ relation. However, it was also mentioned that NGC 4395 could compensate for the deviation with a high accretion rate and might be consistent with the $R_{\mathrm{BLR}}-L$ relation. However, the accretion rate of NGC 4395 is measured to be smaller $\sim$ 0.06. It was stated by \citet{Pei2017} that low accretion rate sources such as NGC 5548 can deviate from $R_{\mathrm{BLR}}-L$, which could support the small deviation of NGC 4395. Estimations of RM masses of more low-luminous sources are needed to address the issue.
\begin{figure}
    \centering
    \includegraphics[height=8cm, width=8.5cm]{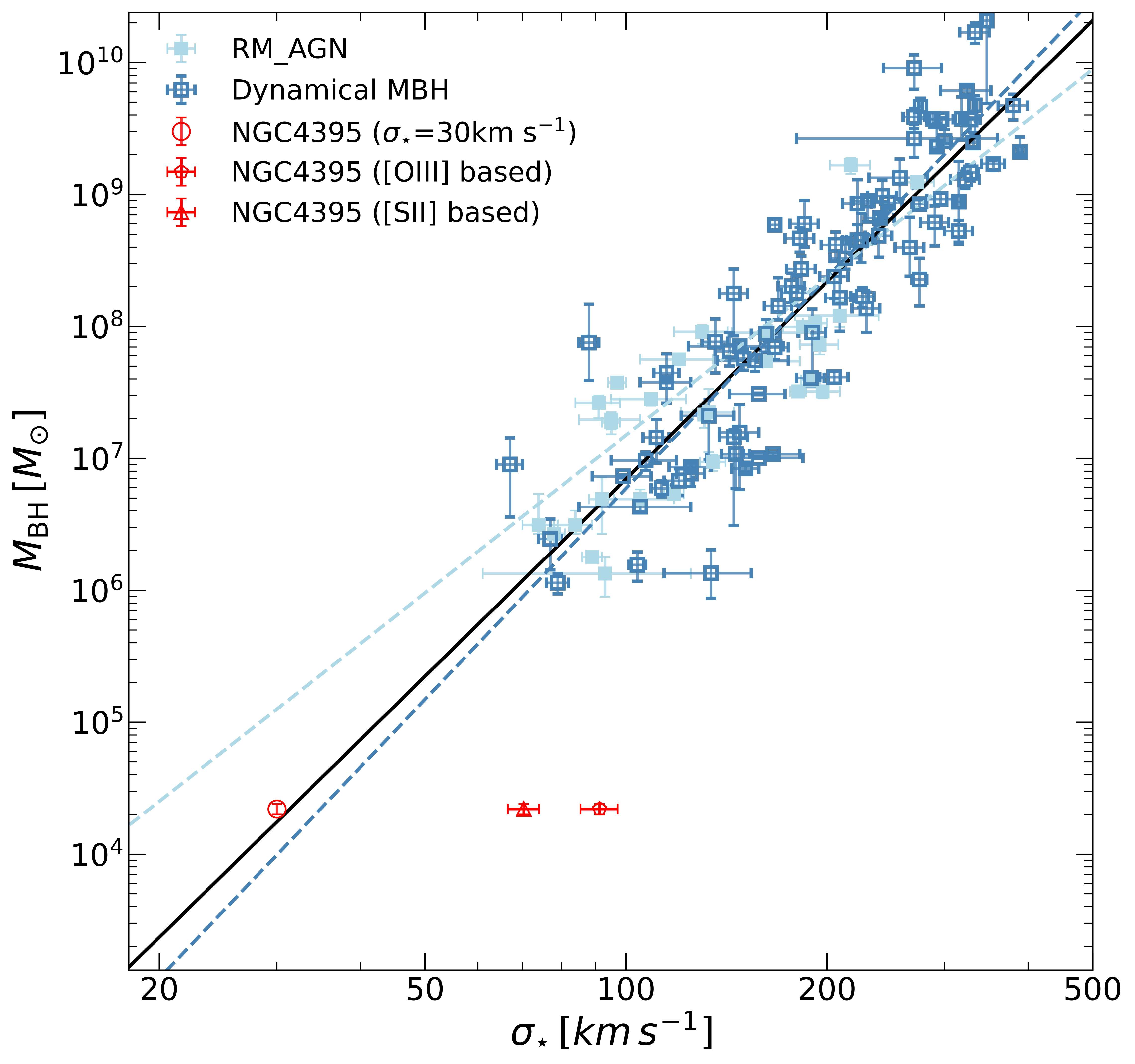}
    \caption{NGC 4395 is plotted on the black hole mass ($M_{\mathrm{BH}}$) vs. stellar velocity dispersion ($\sigma_{\mathrm{\star}}$) diagram. The dynamically measured black hole masses and $\sigma_{\mathrm{\star}}$ for inactive galaxies are in dark-blue open squares taken from \citet{kormendy_coevolution_2013, Woo2015}. The RM-AGN, black hole masses, are shown with light-blue filled squares taken from \citet{Woo2015}. The M$_{\mathrm{BH}}$ for RM-AGNs is based on $\sigma_{\mathrm{line}}$ and the $\sigma_{\mathrm{\star}}$ is also compiled from \citet{Woo2015}. The dark-blue and light-blue dashed lines are the $M_{\mathrm{BH}}$-$\sigma_{\mathrm{\star}}$ fit for dynamical $M_{\mathrm{BH}}$ of inactive galaxies and RM-AGNs, and the solid line illustrating the combined fit for dynamical $M_{\mathrm{BH}}$ and RM-AGN $M_{\mathrm{BH}}$. The slope and intercept are obtained from \citet{kormendy_coevolution_2013} and \citet{Woo2015}, respectively. We plotted the source NGC 4395 with the upper limit of $\sigma_{\mathrm{\star}}$, i.e., 30 km s$^{-1}$ and measured [\ion{O}{3}]$\lambda$5007 and [\ion{S}{2}]$\lambda$6732 dispersion along with our measured $M_{\mathrm{BH}}$= $2.2^{+0.2}_{-0.2}\times 10^{4}M_{\mathrm{\odot}}$.}
    \label{fig:Mass-sigma}
\end{figure}

\subsection{Black hole Mass-stellar velocity dispersion}
Previous studies \citep{Woo2013,kormendy_coevolution_2013, Woo2015} have concluded that the AGNs preserve the same correlation between black hole mass ($M_{\mathrm{BH}}$) with its $\sigma_{\mathrm{\star}}$ as confirmed for inactive galaxies. Hence, can the same correlation be trusted for low-luminous IMBHs? Exploring the lower mass range is crucial to understanding black holes affecting their host galaxy growth and evolution. 

An attempt was made by \citet{brok_measuring_2015} to measure the $\sigma_{\mathrm{\star}}$ of the NSC to constrain the central black hole mass in the NGC 4395 galaxy. Dynamical modeling of molecular hydrogen (H$_{\mathrm{2}}$) gas was employed through imaging and spectroscopy using the Hubble Space Telescope (HST)/Wide Field Camera 3 (WFC3) and Keck II telescope, respectively. Unfortunately, a reliable $\sigma_{\mathrm{\star}}$ could not be obtained due to the absence of CO absorption band heads, typical of evolved stellar populations in nearby galaxies. An upper limit of the velocity dispersion $<$ 30 km s$^{-1}$ \citep[as predicted by][]{filippenko_low-mass_2003} was provided against their unreliable estimation below 15 km s$^{-1}$. Similarly, \citet{Brum2019} found the $\sigma_{\mathrm{\star}}$ of ionized gases such as Pa$\beta$, H$\alpha$, and [\ion{N}{2}] to be around 60 km s$^{-1}$ to the east and 40 km s$^{-1}$ to the west with respect to the nucleus of NGC 4395. However, the molecular hydrogen H$_{\mathrm{2}}$ had $\sigma_{\mathrm{line}}$ $\sim$ 30 km s$^{-1}$ similar to what \citet{filippenko_low-mass_2003} obtained for stellar dispersion velocity ($\sigma_{\mathrm{\star}}$) using Ca II absorption features. The upper limit for $\sigma_{\mathrm{\star}}$ was set to be 30 km s$^{-1}$ \citep[see,][]{filippenko_low-mass_2003} using Ca II absorption features.

Due to the challenges encountered in previous attempts to measure $\sigma_{\mathrm{\star}}$ by previous authors \citep[see,][]{brok_measuring_2015, Brum2019}, the dispersion of narrow emission lines is used as a substitute for $\sigma_{\mathrm{\star}}$. \citet{Sexton2021} and \citet{Le2023} argued that narrow line [\ion{O}{3}]$\lambda$5007 of AGNs is highly correlated with $\sigma_{\mathrm{\star}}$ and hence can be used as a substitute for it. Hence, we measured $\sigma_{\mathrm{[OIII]\lambda5007}}$, which was found to be 91.3$\pm$5.8 km s$^{-1}$ after correcting the instrumental resolution. Additionally, the best-fit single Gaussian modeling of [\ion{S}{2}] line provided a width of 70$\pm$3.8 km s$^{-1}$.

Sources consisting of the mass of inactive galaxies measured using dynamical modeling \citep{kormendy_coevolution_2013} and RM AGNs, compiled by \citet{Woo2015}, are shown in Fig. \ref{fig:Mass-sigma} along with the best-fit relation taken from \citet{kormendy_coevolution_2013} and \citet{Woo2015}. NGC 4395 is placed with previous and current estimates of $\sigma_{\mathrm{\star}}$ with our measured black hole mass, i.e., $2.2^{+0.2}_{-0.2}\times 10^{4}M_{\mathrm{\odot}}$. The source closely follows the $M_{\mathrm{BH}} - \sigma_{*}$ relation if $\sigma_{*}=30$ km s$^{-1}$ is used, however, shows significant offsets when $\sigma_{\mathrm{[OIII]\lambda5007}}$ is used as a surrogate of $\sigma_{\mathrm{\star}}$. We note that \citet{woo2019} found their [\ion{S}{2}] doublets to be better fitted with double Gaussian model for each line, one for core and another for the wing with values $\sigma_{\mathrm{core}}$=18$\pm$1 km $s^{-1}$ and $\sigma_{\mathrm{wing}}$=100$\pm$5 km $s^{-1}$. However, a single Gaussian was enough to model the [\ion{S}{2}] line in our case. Consequently, considering the upper limit of stellar dispersion velocity of 30 km $s^{-1}$ \citep{filippenko_low-mass_2003}, \citet{woo2019} used their measured $\sigma_{\mathrm{core}}$ component in their analysis.

\section{Conclusions}
\label{sec:Conclusions}
The photometric and spectroscopic monitoring were performed for NGC 4395 to measure the BLR size and black hole mass. The main conclusions are:
\begin{enumerate}
    \item The fractional variability for merged V-band and spectroscopic optical continuum flux for 10$^{\mathrm{th}}$ March is 2.6 $\%$, for 11$^{\mathrm{th}}$ March is 7$\%$, and for the entire light curve including both days is 6$\%$ with $R_{\mathrm{max}}$ ranging in 1.12 to 1.30. For the H$\alpha$ light curve, the $F_{\mathrm{var}}$ is maximum at 6.3 \% for the entire light curve with $R_{\mathrm{max}}$ as 1.33.
    
    \item The measured H$\alpha$ BLR size is $125.0^{+6.2}_{-6.1}$ light minutes. This is the best measurement using the second part of the 11$^{\mathrm{th}}$ March light curve.
    
    \item The line width, FWHM and $\sigma_{\mathrm{line}}$, are measured from the mean spectrum constructed from 11$^{\mathrm{th}}$ March second part consisting a total of 24 spectra. The $\sigma_{\mathrm{line}}$ is calculated to be $544.7^{+22.4}_{-25.1}$ km s$^{-1}$ and FWHM 810.2$^{+86.8}_{-91.8}$ km s$^{-1}$.
    
    \item The $\sigma_{\mathrm{line}}$ is used to calculate the black hole mass, which is found to be $2.2^{+0.2}_{-0.2}\times 10^{4}M_{\mathrm{\odot}}$, which lies within the range of masses provided in the literature.
    
    \item The bolometric luminosity (L$_{\mathrm{BOL}}$) is measured as 1.67 $\times$ 10$^{41}$ erg s$^{-1}$ and the Eddington ratio ($\lambda_{\mathrm{EDD}}$) is 0.06.
    
    \item The NGC 4395 is placed in the $R_{\mathrm{BLR}}-L$ relation for SMBHs, where the source shows three times smaller BLR size than expected. RM study of more such low luminous AGN will allow us to calibrate better $R_{\mathrm{BLR}}-L$ relation.
    
    \item Considering the lower limit of $\sigma_{\mathrm{\star}}=30$ km s$^{-1}$ as provided by literature, the source closely follows the $M_{\mathrm{BH}}-\sigma_{\mathrm{\star}}$ relation; however, it shows significant deviation when [\ion{O}{3}]$\lambda$5007 is used as a proxy of $\sigma_{\mathrm{\star}}$.
\end{enumerate}

\begin{acknowledgments}
SR acknowledges the partial support of SERB-DST, New Delhi, through SRG grant no. SRG/2021/001334. KC acknowledges the support of SERB-DST, New Delhi, for funding under the National Post Doctoral Fellowship Scheme through grant no. PDF/2023/004071. This work is supported by the National Research Foundation of Korea (NRF) grant funded by the Korean government (MEST) (No. 2021R1A2C3008486). AKM acknowledges the support from the European Research Council (ERC) under the European Union’s Horizon 2020 research and innovation programme (grant agreement No. [951549]). The study utilizes data from the 3.6-m Devasthal Optical Telescope (DOT), a National Facility, and the 1.3-m Devasthal Optical Telescope (DFOT). Both telescopes are operated and overseen by the Aryabhatta Research Institute of Observational Sciences (ARIES), an autonomous Institute under the Department of Science and Technology, Government of India. Gratitude is extended to the scientific and technical personnel at ARIES DOT and DFOT for their invaluable assistance.
\end{acknowledgments}

\software{MAPSPEC \citep{MAPSPEC}, PyQSOFit \citep{GuoPyqsofit, guo_legolasonpyqsofit_2023}, PyI$^{\mathrm{2}}$CCF code \citep{Guo2022}, PYCALI \citep{PyCALI}, PyCCF \citep{Peterson_1998}, JAVELIN \citep{Zu2011, Zu2013}}
\clearpage

\appendix
\restartappendixnumbering
\section{Detrending lightcurve}
\label{sec:appendix}
\begin{figure}[!htp]
    \centering
    \includegraphics[width=1\linewidth]{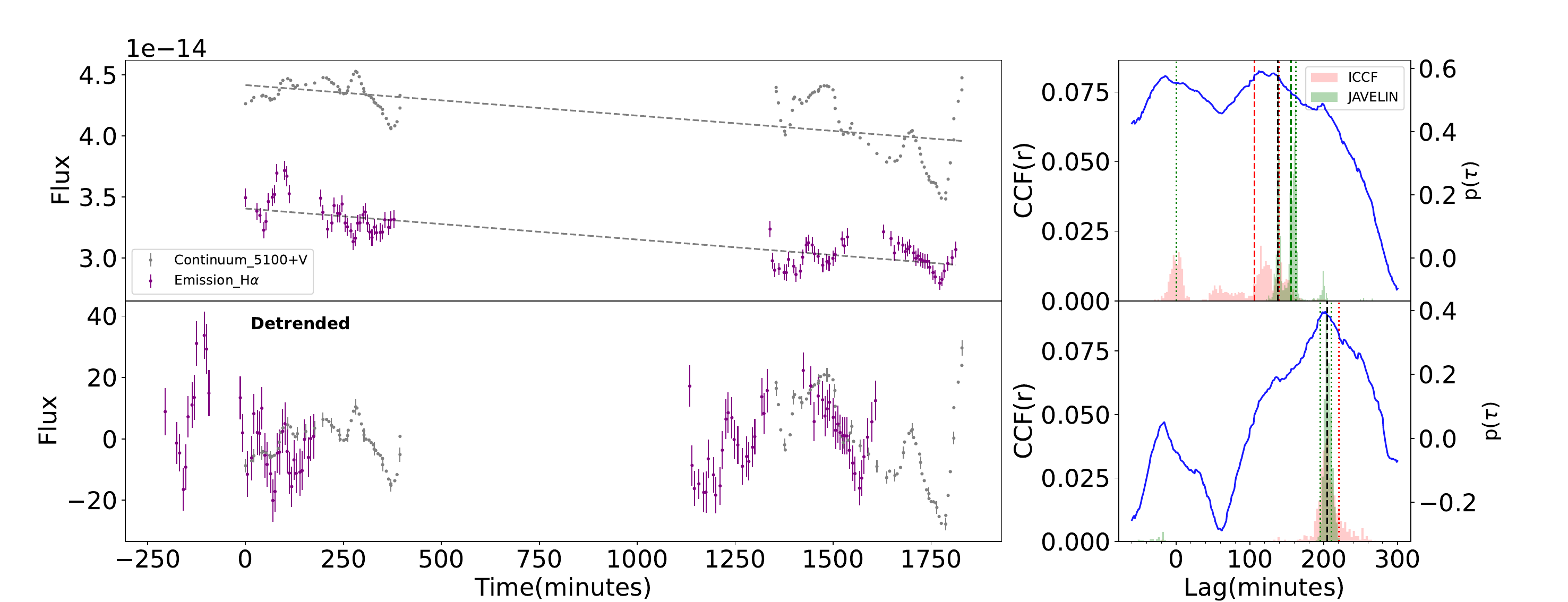}
    \caption{Upper-left panel: The continuum and H$\alpha$ line light curves (arbitrary flux) for the entire two days fitted with a straight line (grey) as mentioned in section \ref{sec:detrended}. Lower-left panel: The mean-subtracted light curves after detrending are shown. Here, the continuum light curve is normalized, and the emission line light curve is shifted by a lag of 205 minutes. Right-panels: The CCF coefficient (r) along with ICCF and JAVELIN lag histograms before (upper-right) and after detrending (lower-right).}
    \label{fig:enter-label}
\end{figure}

We performed detrending of the light curves fitting a straight line in the continuum and line light curve and then subtracted the best-fit straight line to get the detrended light curves as mentioned in section \ref{sec:detrended}. In Figure \ref{fig:enter-label}, we showed the light curve and the lag results before and after detrending for the entire campaign as an example. 

\bibliography{references}{}
\bibliographystyle{aasjournal}
\end{document}